\documentclass[a4paper,11pt]{article}
\usepackage[utf8]{inputenc}
\usepackage{graphicx}
\usepackage{amssymb,amsfonts,amsmath}
\usepackage[authoryear,round]{natbib}
\usepackage{geometry}
\usepackage{longtable}
\usepackage{multirow}
\usepackage{url}

\newcommand{\IntRxn}{\mathcal{R}^M}
\newcommand{\ERxn}{\mathcal{R}^E}
\newcommand{\QRxn}{\mathcal{R}^Q}
\newcommand{\ExcRxn}{\mathcal{R}^T}
\newcommand{\IntMet}{\mathcal{M}}
\newcommand{\Enz}{\mathcal{E}}
\newcommand{\rib}{R}
\newcommand{\Quota}{\mathcal{Q}}
\newcommand{\glyc}{G}
\renewcommand{\v}{\mathbf{v}}
\renewcommand{\c}{\mathbf{c}}
\newcommand{\Irr}{Irr}
\newcommand{\kcat}{\kappa}

\title{Evaluating the stoichiometric and energetic constraints of cyanobacterial diurnal growth} 

\author{Alexandra-M. Reimers\footnote{Department of Mathematics and Computer Science, Freie Universit\"{a}t Berlin, Arnimallee 6, 14195 Berlin, Germany} \footnote{International Max Planck Research School for Computational Biology and Scientific Computing, Max-Planck-Institut f\"ur molekulare Genetik, Ihnestra\ss e 63-73, 14195 Berlin, Germany} \and Henning Knoop\footnote{Fachinstitut Theoretische Biologie (ITB), Institut f\"ur Biologie, Humboldt-Universit\"at zu Berlin, Invalidenstra\ss e 43, 10115 Berlin, Germany} \and Alexander Bockmayr$^*$\and Ralf Steuer$^\ddag$}

\begin{document}

\maketitle

\begin{abstract}
Cyanobacteria are an integral part of the Earth's biogeochemical cycles and a promising resource for the synthesis of renewable bioproducts from atmospheric CO$_2$. Growth and metabolism of cyanobacteria are inherently tied to the diurnal rhythm of light availability. As yet, however, insight into the stoichiometric and energetic constraints of cyanobacterial diurnal growth is limited. Here, we develop a computational platform to evaluate the optimality of diurnal phototrophic growth using a high-quality genome-scale metabolic reconstruction of the cyanobacterium {\it Synechococcus elongatus} PCC 7942. We formulate phototrophic growth as a self-consistent autocatalytic process and evaluate the resulting time-dependent resource allocation problem using constraint-based analysis. 
Based on a narrow and well defined set of parameters, 
our approach results in an {\it ab initio} prediction of growth properties over a full diurnal cycle.
In particular, our approach allows us to study the optimality of metabolite partitioning during diurnal growth. 
The cyclic pattern of glycogen accumulation, an
 emergent property of the model, has timing characteristics that are 
shown to be a trade-off between conflicting cellular objectives.
The approach presented here provides insight into the time-dependent resource allocation problem of phototrophic diurnal growth and may serve as a general framework to evaluate the optimality of metabolic strategies that evolved in photosynthetic organisms under diurnal conditions. 
\end{abstract}

\newpage
Cyanobacterial photoautotrophic growth requires a highly coordinated distribution of cellular resources to different intracellular processes, including the {\it de novo} synthesis of proteins, ribosomes, lipids, as well as other cellular components. For unicellular organisms, the optimal allocation of limiting resources is a key determinant of evolutionary fitness in almost all environments. Owing to the importance of cellular resource allocation for understanding cellular trade-offs, as well as its importance for the effective design of synthetic properties, the cellular 'economy' and its implication for bacterial growth laws have been studied extensively \citep{molenaar2009,scott2010,flamholz2013,vazquez2014,Hui2015,Burnap2015,weisse2015} -- albeit almost exclusively for heterotrophic organisms under stationary environmental conditions. For photoautotrophic organisms, including cyanobacteria, growth-dependent resource allocation is further subject to diurnal light-dark (LD) cycles that partition cellular metabolism into distinct phases. 
Recent experimental results have demonstrated the relevance of time-specific synthesis for cellular growth~\citep{diamond2015}. Nonetheless the implications and consequences of growth in a diurnal environment on the cellular resource allocation problem are insufficiently understood, and computational approaches hitherto developed for heterotrophic growth are not straightforwardly applicable to phototrophic diurnal growth.

Here, we propose a computational framework to evaluate the optimality of diurnal resource allocation for diurnal phototrophic growth. We are primarily interested in the stoichiometric and energetic constraints that shape the cellular 'protein economy', that is, the relationship between the average growth rate and the relative partitioning of metabolic, photosynthetic, and ribosomal proteins during a full diurnal period.  Beyond the established constraint-based reconstruction and analysis methodologies, we aim to obtain an {\it ab initio} prediction of emergent properties that arise from a narrow and well-defined set of assumptions and parameters about cyanobacterial diurnal growth -- and to contrast these emergent properties with known and observed cellular behavior. To this end, we assemble and evaluate an auto-catalytic genome-scale model of cyanobacterial growth, based on a high-quality metabolic reconstruction of the cyanobacterium {\it Synechococcus elongatus} PCC 7942. Our evaluation significantly improves upon a previous model of diurnal cyanobacterial growth~\citep{ruegen2015} and takes into account recent developments in constraint-based analysis~\citep{King2015,Henson2015}. Our approach is closely related to resource balance analysis (RBA)~\citep{goelzer2011}, dynamic enzyme-cost flux balance analysis (deFBA)~\citep{waldherr2015}, as well as integrated metabolism and gene expression (ME) models~\citep{obrien2013}, but explicitly accounts for diurnal phototrophic growth. 

\begin{figure*}[th]
\centering
\includegraphics[width=0.95\textwidth]{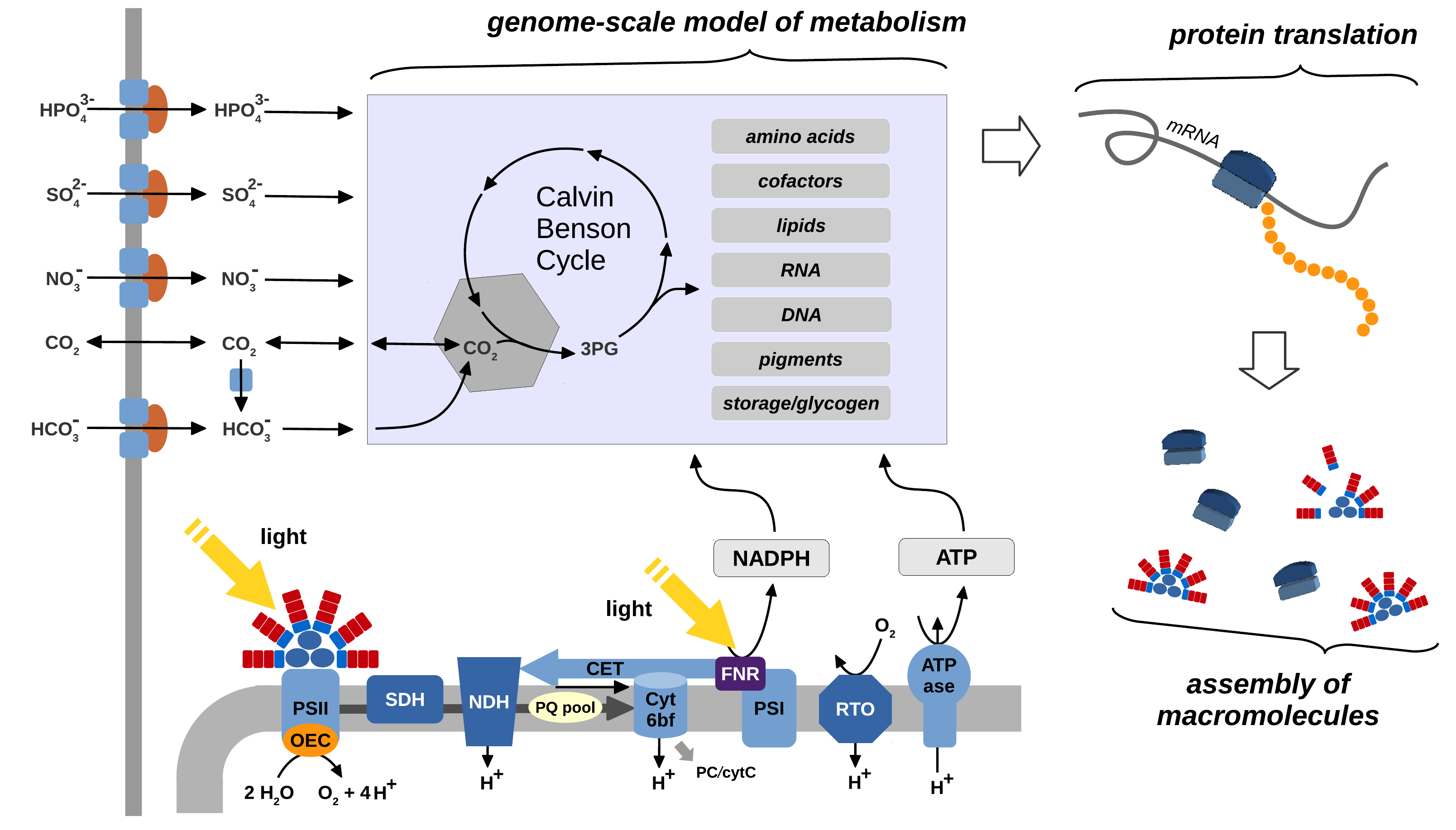}
\caption{A self-consistent autocatalytic growth model of {\it Synechococcus elongatus} PCC 7942. Energy and reducing agents are derived from the photosynthetic light reactions and drive the fixation of inorganic carbon via the Calvin-Benson cycle, as well as the subsequent synthesis of cellular macromolecules. The synthesis of macromolecules is modeled using a genome-scale metabolic reconstruction. The capacity of each metabolic reaction depends on the availability of the respective catalyzing enzymes. Enzymes are translated using their constituent amino acids, which are themselves the products of metabolism. The abundances of all macromolecules relevant to cellular growth (metabolic enzymes, transporters, photosynthetic and respiratory protein complexes, phycobilisomes, and ribosomes) are time-dependent quantities that are governed by the respective differential mass-balance equations. The timing and amount of synthesis of macromolecules constitutes a global diurnal resource allocation problem. 
\label{fig:Model}}
\end{figure*}

Using {\it Synechococcus elongatus} PCC 7942 as a model system, our starting point is the observation that almost all cellular processes are dependent upon the presence of catalytic compounds, typically enzymes and other cellular macromolecules. Hence, a self-consistent description of cyanobacterial growth must take the synthesis of these macromolecules into account -- and reflect the fact that the abundance of these macromolecules limits the capacity of cellular metabolism at all times. {\it De novo} synthesis of cellular macromolecules increases the metabolic capacity -- the timing and amount of the respective synthesis reactions can therefore be described as a cellular resource allocation problem: What is the amount and temporal order of synthesis reactions to allow for maximal growth of a cyanobacterial cell in a diurnal environment? To evaluate the respective stoichiometric and energetic constraints, we only require knowledge about the stoichiometric composition, and the catalytic efficiency of macromolecules -- quantities for which reasonable estimates are available. We therefore seek to evaluate the emergent properties of phototrophic diurnal growth, based on best {\it a priori} estimates of relevant parameters only. Our key results include (i) a prediction of the timing of intracellular synthesis reactions that is in good agreement with known facts about metabolite partitioning during diurnal growth, (ii) limits on the estimated maximal rate of phototrophic growth that are close to observed experimental values, suggesting a highly optimized metabolism, (iii) a predicted optimal timing of glycogen accumulation that is in good agreement with recent experimental findings.



\section*{Results}

\subsection*{Network reconstruction and constraints}
We assembled a model of a self-replicating cyanobacterial cell based on a genome-scale metabolic reconstruction of the cyanobacterium {\it Synechococcus elongatus} PCC 7942. The model incorporates a manually curated representation of all key processes relevant to the energetics of phototrophic growth: Photons are absorbed by light-harvesting antennae, the phycobilisomes, attached primarily to photosystem II (PSII). The energy derived from absorbed photons drives water splitting at the oxygen-evolving complex (OEC) and, via the photosynthetic electron transport chain (ETC), results in the regeneration of cellular ATP and NADPH. The ETC consists of a set of large protein complexes, PSII, cytochrome b$_6$f complex (Cytb$_6$f), photosystem I (PSI), and ATP synthase (ATPase), embedded within the thylakoid membrane. Inorganic carbon is taken up via CO$_\mathrm{2}$ concentrating mechanisms (CCMs) and assimilated via the Calvin-Benson cycle. The product of the ribulose-1,5-bisphosphate carboxylase/oxygenase (RuBisCO), 3-phosphoglycerate (3PG), serves as a substrate for the biosynthesis of cellular components, such as DNA, RNA, lipids, pigments, glycogen, and amino acids. Cellular metabolism is represented by a detailed genome-scale reconstruction of {\it Synechococcus elongatus} PCC 7942. Amino acids serve as building blocks for structural, metabolic, photosynthetic, and ribosomal proteins. All cellular components are represented by their known molecular composition. The model is depicted in Figure~\ref{fig:Model} and detailed in the Materials and Methods. It encompasses a total of $465$ macromolecules and $1112$ reactions, including $645$ metabolic and exchange reactions, $616$ metabolic genes, as well as $467$ compound production reactions.

\subsection*{Phototrophic growth is autocatalytic}
To implement the conditional dependencies of phototrophic growth, the rate of each process is constrained by the abundances of the respective catalyzing macromolecules and their respective catalytic efficiencies. For example, at any point in time, each individual metabolic reaction is constrained by the abundance of its catalyzing enzyme (or enzyme complex) and the respective catalytic turnover number~$k_\mathrm{cat}$. The latter values are globally sourced from databases \citep{brenda,sabiork}, see Materials and Methods. Protein synthesis is limited by the abundance of ribosomes and modeled according to general principles of peptide elongation, taking into account energy expenditure (one ATP and two GTPs per amino acid) and coupling to metabolism. Light absorption at PSII is constrained by the reported effective cross-section of phycobilisomes and depends on (variable) phycobilisome rod length. Detachment of phycobilisomes from PSII reduces energy transfer to the OEC. For simplicity, light absorption at PSI is assumed to take place in the absence of phycobilisomes using an effective cross-section per PSI complex and energy spillover from PSII is not considered (see Supplementary Text for further discussion). For the photosynthetic and respiratory electron transport chains, maximal catalytic rates per protein complex are sourced from the literature. We note that all aforementioned dependencies only constrain maximal rates of processes, actual rates may be lower due to (unknown) fractional saturation of reaction rates. 

\subsection*{Dynamic resource allocation}~During a full LD cycle, the capacity constraint induced by the abundance of catalyzing compounds on each maximal reaction rate must be fulfilled at each point in time. Catalyzing compounds, however, can be synthesized {\it de novo}, depending on available resources, and may therefore accumulate over a diurnal period, and hence increase the capacity of the respective reactions. To this end, the abundances of macromolecules (metabolic enzymes, transporters, photosynthetic and respiratory protein complexes, phycobilisomes, and ribosomes) are time-dependent quantities that are governed by the respective differential mass-balance equations. To solve the global resource allocation problem, the mass-balance equations including the abundance-dependent rate constraints are cast into a linear programming (LP) problem. The LP-problem is supplemented by periodic boundary conditions for the macromolecules of the form 
\begin{equation} \label{eq:growth}
M(t_0+ 24\mathrm{h}) = \mu \cdot M(t_0) ~~,
\end{equation}
where $M(t)$ denotes (absolute) abundances of time-dependent cellular components at time~$t$, $t_0$ is the initial time, and $\mu$ the multiplication factor. The elements of $M$ at time $t$ are themselves an outcome of the resource allocation problem and not specified externally. Time is discretized using a Gau\ss{} implicit method (midpoint rule). We are primarily interested in diurnal dynamics, and hence a time-scale of several hours. Following the arguments of R\"{u}gen et al.~\citep{ruegen2015} and Waldherr et al.~\citep{waldherr2015}, we therefore assume that internal metabolites are in quasi-steady-state. Equation~[\ref{eq:growth}] represents balanced growth in a periodic environment. Specifically, we assume stationary diurnal experimental conditions, such that the average measured cellular composition per unit biomass after a full diurnal period is invariant. Equation~[\ref{eq:growth}], in conjunction with the mass-balance constraints, the abundance-dependent rate constraints, and the growth objective, $\mu \longrightarrow \mathrm{max}$, define a self-consistent resource allocation problem for diurnal phototrophic growth of the cyanobacterium {\it Synechococcus elongatus} PCC 7942. As input parameters, we require the stoichiometric composition of macromolecules in terms of their constituent amino acids and micro-nutrients, as well as their catalytic efficiencies per enzyme or enzyme-complex. We argue that reasonable approximation of both quantities exist for almost all cellular macromolecules. Using this narrow and well-defined set of parameters, we seek to derive the emergent properties of diurnal phototrophic growth without making use of any further ad-hoc assumptions about metabolic functioning or regulation. For details of the implementation and a discussion of the limits of applicability see the Materials and Methods and the supplement.

\begin{figure}[t]
\centering{\includegraphics[width=.70\textwidth]{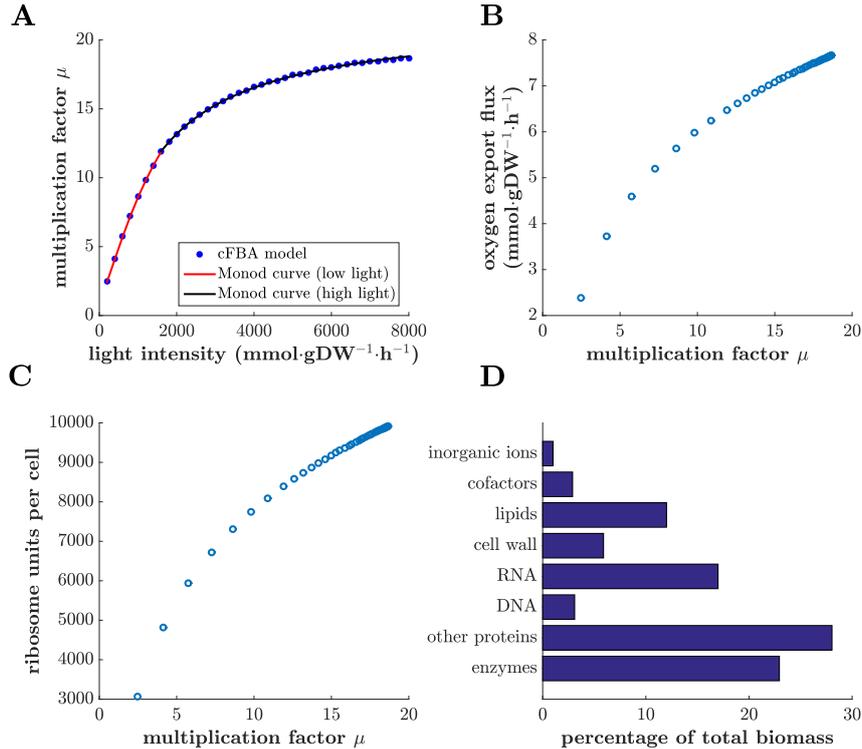}}
\caption{Balanced growth under constant light. 
(A) The maximal multiplication factor $\mu$ as a function of light intensity. Light uptake depends on the maximal effective cross sections of the photosystems. Simulation results indicate that the literature-sourced value $\sigma_\mathrm{PSII} \approx 1\, \mathrm{nm}^2$~\citep{Mackenzie2004} underestimates the actual effective cross section. 
The resulting growth rate $\lambda$ as a function of light intensity
is consistent with the Monod equation (fitted curves are converted into the corresponding multiplication factor $\alpha$). 
(B) Oxygen (O$_\mathrm{2}$) evolution as a function of $\mu$. (C) Ribosome content per cell as a function of $\mu$, assuming a cell mass of $1.5 \, \mathrm{pg}$. (D) The emergent cellular composition for a light intensity of $200~\mu\mathrm{mol}$ photons $\cdot \mathrm{m}^{-2}\cdot\mathrm{s}^{-1}$. 
\label{fig:constLight}}
\end{figure}

\subsection*{Growth under constant light}~Prior to the evaluation of diurnal dynamics, we evaluate light-limited growth under constant light. The uptake of all other nutrients, in particular inorganic carbon, is described by simple Michaelis-Menten uptake reactions and only constrained by the availability of the respective transporters. Carbon cycling is not considered explicitly, the respective energy expenditure is considered as part of general maintenance. Solving the global resource allocation problem, we obtain the multiplication factor $\mu$ and the growth rate $\lambda = \log(\mu)/(24 \mathrm{h})$ as a function of light intensity, as well as the cellular composition for different growth rates. Key results are shown in Figure~\ref{fig:constLight}. For comparison with conventional flux balance analysis (FBA), we use a light intensity $I = 150~\,\mathrm \mu\mathrm{mol}$ photons$\; \mathrm{ s^{-1}\;m^{-2}}$, resulting in the absorption of $15.9 \, \mathrm{mmol}$ photons$\;\mathrm{gDW^{-1}\; h^{-1}}$, a growth rate of $\lambda = 0.03~\mathrm{h^{-1}}$ (multiplication factor $\mu \approx 2$), and an oxygen evolution rate of $1.92 \, \mathrm{mmol\; gDW^{-1} \; h^{-1}}$. These values are in excellent agreement with previous estimates using FBA~\citep{Knoop2013}, and the respective experimental data~\citep{young2011}.
In particular, evaluating the metabolic reconstruction of {\it Synechococcus elongatus} PCC 7942 with conventional FBA and a static biomass objective function (BOF) using a light uptake of $15.8 \, \mathrm{mmol}$ photons $\mathrm{gDW^{-1} \, h^{-1}}$ absorbed, results in an oxygen evolution rate of $1.92 \, \mathrm{mmol\; gDW^{-1} \; h^{-1}}$ and a growth rate of $\lambda = 0.03~\mathrm{h^{-1}}$. In contrast to the static BOF used in FBA, the cellular composition of the autocatalytic model is an emergent result of the global resource allocation problem (Figure~\ref{fig:constLight}D), and is in good agreement with  previously reported BOFs~\citep{nogales2012,Knoop2013}.

When evaluating different light intensities, the growth rate and oxygen evolution increase with increasing light (Figure~\ref{fig:constLight}A and~\ref{fig:constLight}B). We note that light uptake depends on the assumed maximal effective cross section of PSII, reported to be  $\sigma_\mathrm{PSII} \approx 1\, \mathrm{nm}^2$~\citep{Mackenzie2004} -- the results shown in Figure~\ref{fig:constLight} indicate that the reported value underestimates the actual effective cross section (with no further impact on model results, see also Supplementary Text). Similar to findings for models of heterotrophic growth, the relative amount of ribosomes increases with increasing growth rate (Figure~\ref{fig:constLight}C). We observe that growth as a function of light saturates at a growth rate of $\lambda_\mathrm{max} = 0.1281~\mathrm{h^{-1}}$ (multiplication factor $\mu \approx 18$),
estimated using a Monod growth equation (Supplementary Figure S2).
The maximal growth rate is slightly slower than the maximal growth rate observed for {\it Synechococcus elongatus} PCC 7942, reported as $\lambda = 0.14~\mathrm{h^{-1}}$ by Yu et al.~\citep{Yu2015}. We therefore performed a sensitivity analysis of growth rate as a function of estimated parameters. While the sensitivity with respect to the catalytic efficiencies of invidual enzymes is rather low (Supplemental Figures S3-S5), a  major determinant of maximal growth rate is the assumed ratio of non-catalytic (quota) proteins (Supplemental Figure S6). Based on recent proteomics data for slow growing cells~\citep{guerreiro2014}, the ratio was determined to be $55$\% of total protein. No experimental estimates exist for fast growing cells. If the actual percentage for fast growing cells is assumed to be $\sim 20$\%, the resulting growth rates ($\lambda_\mathrm{max} \approx 0.20~\mathrm{h^{-1}}$,  corresponding to a division time of $T_\mathrm{D} =3.5 \mathrm{h}$) are in good agreement and slightly exceed fastest known growth rates of {\it Synechococcus elongatus} PCC 7942~\citep{Yu2015}. 

\begin{figure}
\centering{ \includegraphics[width=0.45\textwidth]{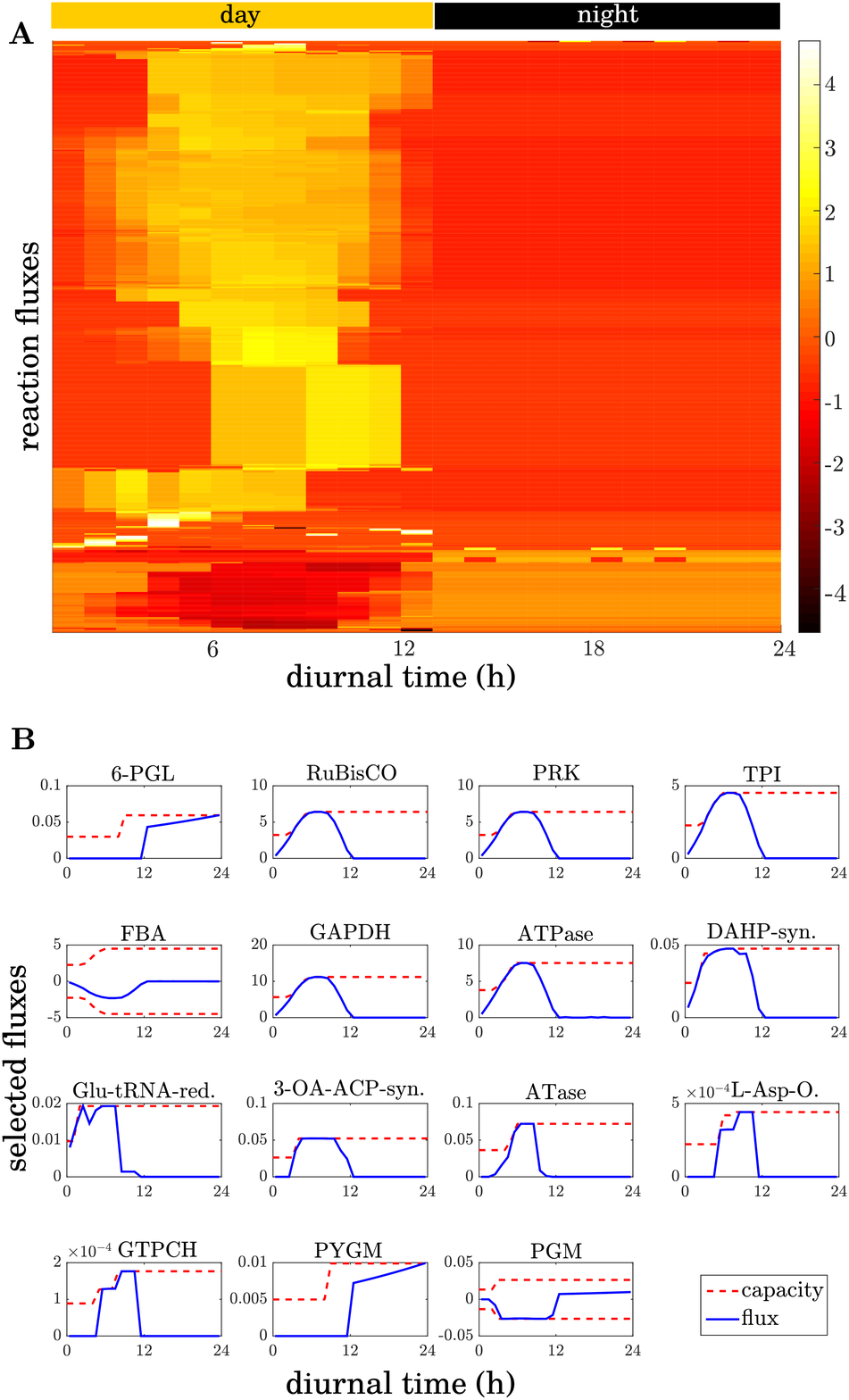}}
\caption{The solution of the time-dependent resource allocation problem over a full diurnal cycle. (A) Metabolic fluxes as a function of time. Cellular metabolism is organized into distinct temporal phases, ranging from synthesis of amino acids and pigments, to synthesis of lipids, DNA/RNA and peptidoglycan, to synthesis of co-factors. (B) Selected metabolic fluxes. Dashed red lines indicate the respective enzymatic capacities (proportional to enzyme amount). Negative values indicate reversible reactions. Photosynthesis and reactions of central metabolism closely follow light availability, as exemplified by the reactions of the Calvin-Benson cycle (RuBisCO, PRK, TPI, FBA, GAPDH) and ATPase. Metabolic activity during the early light period also includes amino acid synthesis (e.g. DAHP-syn.) and synthesis of pigments (e.g. Glu-tRNA-red. towards chlorophyll). Later, metabolic activity shifts to lipid synthesis (e.g. 3-OA-ACP-syn.), followed by synthesis of co-factors (such as L-Asp-O. towards nicotinamide adenine dinucleotide or GTPCH towards tetrahydrofolate (THF)). Glycogen is accumulated during the day, involving flux via the phosphoglucomutase (PGM). During night, glycogen is utilized via the glycogen phosphorylase (PYGM) and serves as a substrate for the pentose phosphate pathway (e.g. 6-phosphogluconolactonase, 6-PGL). We note that synthesis of enzymes can significantly precede reaction flux (see e.g. PYGM). {\it Abbreviations:} RubisCO, ribulose-1,5-bisphosphate carboxylase/oxygenase; PRK, phosphoribulokinase; TPI, triosephosphate isomerase; FBA, fructose-bisphosphate aldolase; GAPDH, glyceraldehyde 3-phosphate dehydrogenase; DAHP-syn., 3-deoxy-D-arabinoheptulosonate 7-phosphate synthase; Glu-tRNA-red., glutamyl-tRNA reductase; 3-OA-ACP-syn., 3-oxoacyl-acyl-carrier-protein synthase; ATase, amidophosphoribosyltransferase; L-Asp-O., L-Aspartate oxidase; GTPCH, GTP cyclohydrolase I.\label{fig:ReferenceDay}}
\end{figure}

\subsection*{A day in the life of Synechococcus elongatus 7942}
Going beyond constant light conditions, we next evaluate the global
resource allocation problem for diurnal light conditions as a dynamic
optimization problem with the objective $\mu \longrightarrow \max$. After discretization, the problem is transformed into a sequence of linear optimization problems and solved to global optimality using a binary search.
%
%
We emphasize that our approach does not impose any constraints on the timing of specific synthesis reactions. Rather, the resulting time-courses as well as the cellular composition $M(t)$ are emergent properties of the global resource allocation problem. The light intensity was modeled as a sinusoidal half-wave with a peak light intensity of $600 \, \mu\mathrm{mol}$ photons $\mathrm{ s^{-1}\;m^{-2}}$. Figure~\ref{fig:ReferenceDay}A shows the resulting flux values for a reference day as a function of diurnal time. We observe that most metabolic activity takes place during the light period. In the absence of light, glycogen is mobilized and utilized for cellular maintenance, serving as a substrate for cellular respiration via the pentose phosphate pathway and ultimately cytochrome C oxidase. Figure~\ref{fig:ReferenceDay}B shows selected metabolic fluxes as a function of time together with the respective enzymatic capacity (dashed lines). The observed flux activity is in good agreement with known facts about metabolite partitioning during diurnal growth~\citep{diamond2015}: In the presence of light, carbon is imported and assimilated via the Calvin-Benson cycle. Carbon assimilation and photosynthesis follow light availability. Synthesis of macromolecules is distributed over the light period (Figure~\ref{fig:ReferenceDay}B). Firstly, at dawn, fluxes related to central metabolism, amino acids and pigment synthesis increase. Secondly, reactions with respect to lipid synthesis, DNA/RNA synthesis, and peptidoglycan synthesis exhibit increased flux. Finally, reactions related to {\it de novo} synthesis of co-factors (NADPH, THF, TPP, FAD) carry flux. At dusk, almost all metabolic activities cease. Dark metabolism is dominated by utilization of storage products and respiratory activity: Stored glycogen is mobilized and consumed via the oxidative pentose phosphate pathway (OPPP), thereby generating NADPH for the respiratory electron transport chain. The global cellular resource allocation problem gives rise to a highly coordinated metabolic activity over a diurnal period. The numerical results are highly robust with respect to changes in parameters. Growth rates and overall cellular composition (Supplementary Figure S7) depend on peak light intensity, the results ({Supplementary Figures S8 and S9}) are qualitatively similar to the case of constant light shown in Figure~\ref{fig:constLight}.

\subsection*{Glycogen dynamics for variable day length}
\begin{figure}
\centering{ \includegraphics[width=0.70\textwidth]{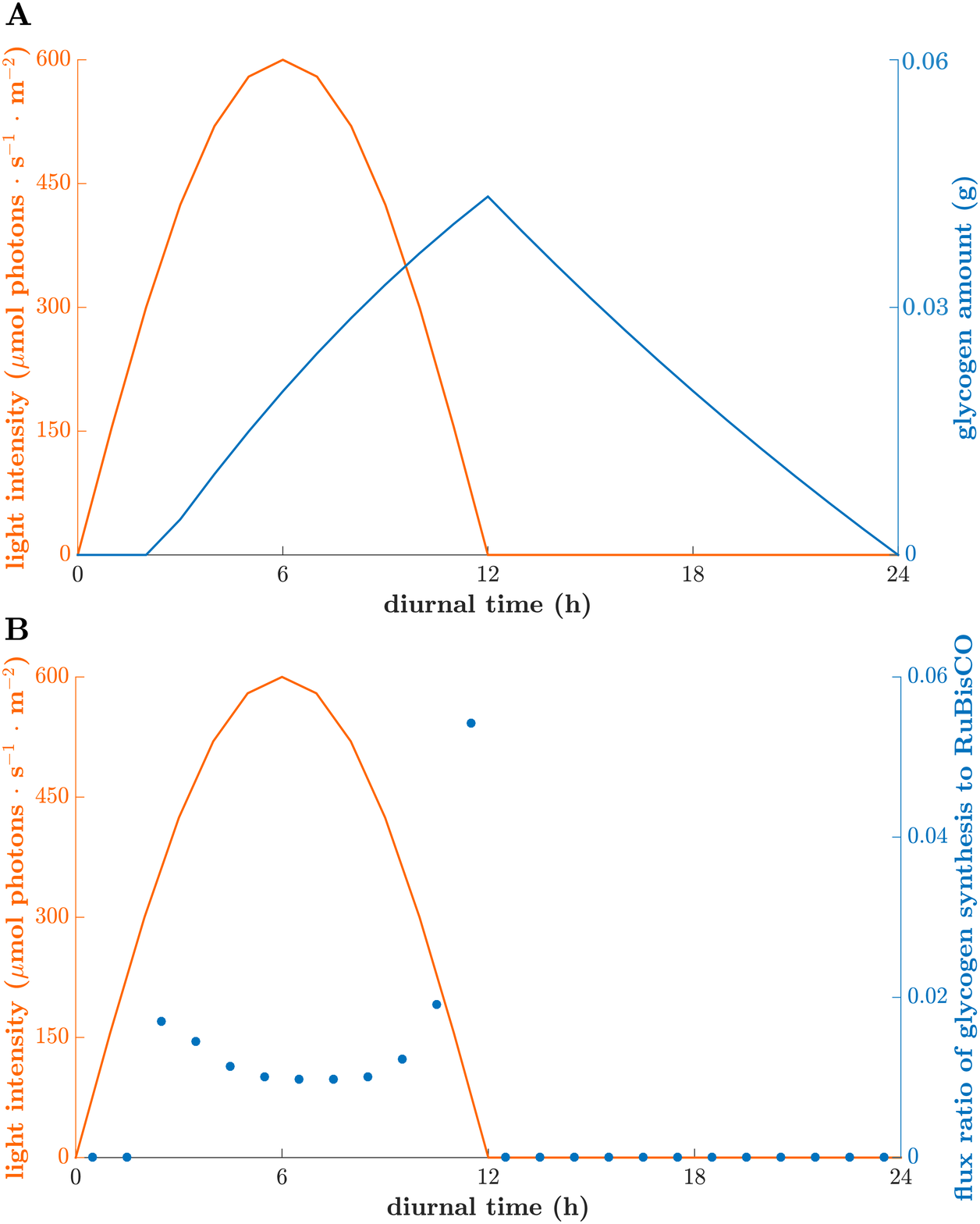}}
\caption{Timing and dynamics of glycogen accumulation over a full diurnal cycle. (A) Cells accumulate glycogen during the light phase and mobilize it as a source of carbon and energy during the night. The model makes no assumptions about timing and amount of glycogen accumulation. Rather, glycogen accumulation emerges as a trade-off between conflicting objectives. Shown are absolute amounts of stored glycogen per gDW. (B) Linear accumulation of glycogen requires control of the carbon partitioning ratio. Shown is the ratio of glycogen synthesis with respect to carbon fixation (per carbon). \label{fig:Glycogen}}
\end{figure}

\begin{figure}
\centering{ \includegraphics[width=0.50\textwidth]{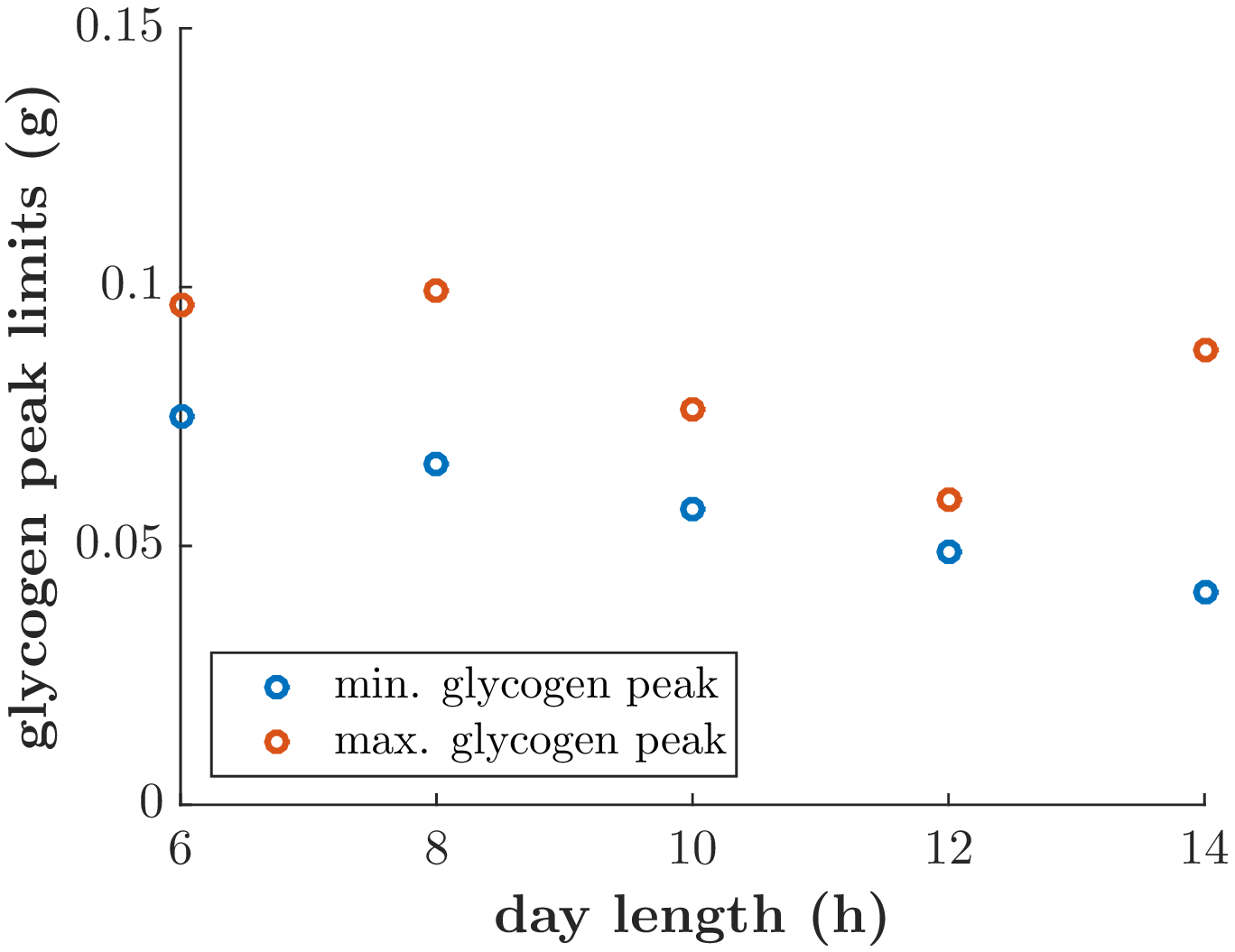}}
\caption{Minimal glycogen requirements for different length of the light period (day length). Peak glycogen content is always observed at dusk. The absolute amount of glycogen required at dawn exhibits variability, indicating plasticity of metabolism with respect to the resource allocation problem. Higher glycogen at dusk implies increased metabolic activity shortly before dawn at the expense of slightly reduced
synthesis reactions during the light period. 
\label{fig:minGlycogen}}
\end{figure}

~Glycogen is the main storage compound in cyanobacteria. Cells accumulate glycogen during the light phase and mobilize it as a source of carbon and energy during the night. It was recently shown that the timing of glycogen accumulation is under tight control of the cyanobacterial circadian clock and disruption of the clock results in altered glycogen kinetics~\citep{diamond2015}. We therefore evaluate the optimality of glycogen accumulation in the context of the global resource allocation problem. We note that our simulation does not impose any {\it ad hoc} constraints on the kinetics and timing of glycogen synthesis. Rather, accumulation of glycogen is a systemic property that emerges as a consequence of optimal resource allocation. Figure~\ref{fig:Glycogen}A shows the time course of glycogen accumulation obtained from the global resource allocation problem over two diurnal periods. Figure~\ref{fig:Glycogen}B shows the optimal carbon partitioning during the light period. Stored glycogen increases linearly within the light period, in good agreement with recent data from {\it Synechococcus elongatus} 7942~\citep{diamond2015}, and {\it Synechocystis} sp. PCC 6803~\citep{Saha2016}. We note that a linear slope is not self-evident, but emerges as a trade-off between at least two conflicting objectives: minimal withdrawal of carbon during the early growth period (favoring carbon withdrawal later in the day) versus a minimal capacity requirement for the synthesis pathway (favoring constant withdrawal throughout the light period).
To further highlight glycogen accumulation as a systemic property, we evaluate the minimal amount of accumulated glycogen for different light periods. Figure~\ref{fig:minGlycogen} shows the results for different lengths of day versus night periods. If the night period is doubled, slightly less than twice the glycogen is required to sustain night metabolism and the amount of glycogen required at dusk exhibits a certain plasticity.
The latter fact corresponds to differences in resource allocation with no discernible effect on overall growth: Certain synthesis tasks, in particular lipid synthesis, can be relegated to the end of the night period, thereby requiring less enzyme capacity during the day at the expense of an increased glycogen requirement at dusk (Supplementary Figure S11-S12).   


\section*{Discussion}
Phototrophic growth under diurnal conditions requires a precise coordination of metabolic processes -- which is challenging to describe using conventional FBA and related constraint-based approaches~\citep{Henson2015}. In this work, we developed a genome-scale model that allows us to evaluate the stoichiometric and energetic constraints of diurnal phototrophic growth in the context of a global diurnal resource allocation problem. Building upon previous works~\citep{molenaar2009,goelzer2011,ruegen2015,waldherr2015,Burnap2015,obrien2013}, our approach is based on the fact that growth is inherently autocatalytic: The cellular machinery to sustain metabolism is itself a product of metabolism. 
Our focus were the net stoichiometric and energetic implications of diurnal growth on a time-scale of several hours, in particular related to the {\it de novo} synthesis of proteins and other cellular macromolecules. Faster time-scales, in particular a detailed representation of macromolecular assembly, were not considered. We consider our approach to be appropriate for cells with a division time of approximately 24h or faster under diurnal light conditions. For very slow growing cells, the importance of {\it de novo} synthesis of proteins is likely diminished, and other cellular processes become dominant, such as protein turnover, maintenance and repair mechanisms. 

Given these limits of applicability, our aim was an {\it ab initio} prediction of optimal diurnal resource allocation: How is metabolism organized over a full diurnal cycle? How are the synthesis reactions of cellular macromolecules organized over a full diurnal cycle?  What is the optimal timing of glycogen accumulation during the light phase? Importantly, from the perspective of resource allocation, these questions can be evaluated without extensive knowledge of kinetic parameters and regulatory interactions. Our analysis is based solely on knowledge of the stoichiometric compositions and the turnover numbers of catalytic macromolecules -- reasonable estimates for both quantities are available and the respective values were sourced from the primary literature and databases.

Our results, similar to time-independent FBA, are based on the assumption of optimality, and hence allow us to pinpoint the energetic trade-offs and constraints relating to diurnal growth. Overall, the {\it ab initio} results obtained from the global resource allocation problem are in good agreement with previous knowledge and experimental observations about flux partitioning in {\it Synechococcus elongatus} 7942. Growth predominantly takes place during the light phase. In the absence of light, almost all metabolic activity ceases, and cellular metabolism is  dominated by respiratory activity. Carbon fixation and central metabolism largely follow light availability, whereas other synthesis reactions follow a specific temporal pattern -- including synthesis of macromolecules well before their utilization.

We note that cessation of metabolic activity during darkness is itself already a result of a trade-off between idle enzymatic capacity versus the energy requirements for synthesis reactions. As shown, similar to the observation in a previous minimal model 
\citep{ruegen2015}, an {\it in silico} experiment with artificially lowered enzyme costs for glycogen synthesis and mobilization, results in increased utilization of synthesis reactions at night -- thereby minimizing requirements for enzyme capacity at the expense of additional storage capacity.
%
In this respect, the function of glycogen is analogous to a cellular battery or capacitor -- and the timing of glycogen synthesis results as a trade-off between conflicting objectives: early withdrawal of carbon from an auto-catalytic system versus minimizing glycogen synthesis capacity versus extending the time span of enzyme utilization. 

We consider our approach to be a suitable general framework to evaluate the optimality of diurnal phototrophic growth. As a first test, we considered the maximal growth rate, as predicted by optimal resource allocation using independently sourced parameters only. The results show that model-derived values are indeed within the typical range of cyanobacterial growth rates. Since several detrimental factors, such as possible photoinhibition, are not explicitly considered within our model, the close correlation between observed and model-derived growth rates suggests that cyanobacterial metabolism indeed operates close to optimality, in particular when considering high growth rates. In this respect, an unknown factor is the relative amount of non-catalytic (quota) proteins, estimated to be up to 55\% of total protein for slow growing cells~\citep{guerreiro2014}. We conjecture that for fast growing cells this percentage is considerably lower. Indeed, the importance of non-catalytic (quota) proteins as (environment-specific) niche-adaptive proteins (NAPs) on the maximal growth rate was already discussed by Burnap~\citep{Burnap2015}.

In future iterations, our approach can be significantly improved upon.
Of particular interest are the energetic implications of carbon cycling~\citep{Mangan2016} in growing cells, light damage and its repair, as well the temporal coordination of nitrogen fixation in certain cyanobacteria. More generally, we conjecture that the global resource allocation problem described here allows us to evaluate the cost of individual genes and genomes~\citep{Lynch2015} in the context of a growing cell, and thereby allows us to evaluate metabolic adaptations and the diversity of cyanobacterial metabolism~\citep{Beck2012} -- ultimately aiming to understand the limits of phototrophic growth in complex environments.





\section*{Materials and Methods}

\subsection*{Metabolic network model}
All simulations are based on a genome-scale conditional FBA (cFBA) model \citep{ruegen2015}. The model is derived from a genome-scale metabolic reconstruction of the cyanobacterium {\it Synechococcus elongatus} PCC 7942. The reconstruction covers $616$ genes and consists of $662$ metabolic reactions and $539$ metabolites. The reconstruction process was analogous to previous reconstructions~\citep{Knoop2010,Knoop2013}. The original metabolic network reconstruction is provided as Supplementary File 2 (SBML).

\subsection*{Model components and their role}
The cFBA model consists of three types of components: steady-state metabolites, quota components, and components with catalytic function. Quota components have no explicit catalytic function within the model but their synthesis contributes to overall energy and carbon expenditure. We note that, different from ME models~\citep{obrien2013}, we do not aim for a mechanistic representation of processes such as transcription, translation or assembly of macromolecules. Rather, we focus on overall energetic and stoichiometric constraints on diurnal time-scales (several hours).

\subsection*{Components with catalytic activity}
Enzymes, ribosomes and several macromolecules are denoted as components with catalytic function. For each of these components a synthesis reaction is implemented. Macromolecules (e.g. photosystems) assemble once all constituent compounds (amino acids or protein subunits) are available. All components are synthesized using their molecular stoichiometry, as derived from the amino acid sequence. Special attention is paid to the stoichiometries of important photosynthesis and respiration complexes, such as the photosystems or the ATPase. The respective stoichiometries  are listed in the Supplementary Material Tables A1-A8. The amounts of all components with catalytic activity are time-dependent quantities and at each point in time their amount provides an upper limit to the rates of the reactions they catalyze. Assuming that a component (e.g. an enzyme) $e$ catalyzes a reaction $r$, we impose the capacity constraint 
\begin{equation}\label{eq:enzymeCapacity}
v_r(t)\leq M_e(t)\cdot k_{cat_e}^r, \forall t\geq 0
\end{equation}
where $v_r(t)$ denotes the flux through reaction $r$ at time $t$, $M_e(t)$ denotes the concentration of enzyme $e$ at time $t$, and $k_{cat_e}^r$ is the turnover number of the enzyme $e$ for reaction $r$. In case several reactions are catalysed by the same component or enzyme, the sum of their fluxes, weighted by the turnover rates, is bound by the enzyme amount. The capacity constraint holds analogously for all macromolecules, including the components of the electron transport chain and ribosomes.

\subsection*{Quota components}
Main quota components are the vitamins, several cofactors, lipids, cell wall, inorganic ions, DNA, RNA, as well as non-metabolic proteins. These components have to be produced at the same rate as catalytic components, although they do not reinforce the autocatalytic cycle. We enforce their synthesis by imposing an initial amount proportional to their fraction of the whole cell weight and require balanced growth (equation~[\ref{eq:growth}]). Non-metabolic (quota) proteins compete with catalytic proteins for ribosomal capacity. 

\subsection*{Steady-state components}
Turnover of metabolic reactions is considerably faster than the {\it de novo} synthesis of proteins.
Following earlier work~\citep{ruegen2015,waldherr2015}, we therefore assume  internal metabolites to be at quasi-steady-state. The concentrations of internal (non-exchange) metabolites are not explicitly represented in equation~[\ref{eq:growth}] and the metabolic network is assumed to be balanced at all time points. Similar to conventional FBA, we neglect dilution by growth of internal metabolites.  

\subsection*{Turnover rates}
The turnover rates used in the capacity constraint equation~[\ref{eq:enzymeCapacity}] are sourced from the BRENDA database \citep{brenda}. We computationally retrieved all wild type values from all organisms for each enzyme and assigned the median of the corresponding retrieved values as the turnover number of the respective enzyme. For enzymes with no turnover numbers available, we followed~\citep{shlomi2011} and assigned the median of all retrieved turnover numbers. Turnover numbers for the $7$ macromolecules of the ETC were sourced from the primary literature and are listed in Table~\ref{catatable}. The ribosomal capacity is assumed to be $15$ amino acids per second~\citep{young1976}. 

\begin{table}[t]
\centering
\caption{Parameters for the cFBA model. Solving the global resource allocation problem requires knowledge of the catalytic turnover numbers of macromolecules. All values are sourced from the primary literature. }
\begin{tabular}{l r c}
\hline
\textbf{Compound} & \textbf{Catalytic efficiency} & \textbf{Reference}\\
\hline
PSI & 500 s$^{-1}$ & \citep{vermaas2001} \\
PSII & 1000 s$^{-1}$ & \citep{vermaas2001} \\
NDH-1 & 130 s$^{-1}$ & \citep{teicher1998} \\
Cytb6f & 200 s$^{-1}$ & \citep{vermaas2001} \\
Cyt c oxidase & 670 s$^{-1}$ & \citep{howitt1998} \\
SDH & 1300 s$^{-1}$ & \citep{cooley2001} \\
ATPase & 1000 s$^{-1}$ & \citep{nitschmann1986} \\
ribosome & 15 amino acids/s & \citep{young1976} \\
enzymes & various & \citep{brenda} \\
\hline
\end{tabular}
\label{catatable}
\end{table}

\subsection*{Maintenance requirements}
In addition to the processes explicitly included within the model, cells have an additional energy expenditure, usually denoted as {\it maintenance} in FBA models. Along similar lines, our model includes a basal maintenance constraint that hydrolyses ATP with a rate of 0.13 $\mathrm{mmol\; gDW^{-1}\; h^{-1}}$.

\subsection*{Optimization objective and solving routine}
The objective of the resource allocation problem is to maximize the multiplication factor $\mu$ in equation~[\ref{eq:growth}]. The dynamic variables are discretized in time using the implicit midpoint rule numerical scheme. Discretization yields a set of linear constraints, that, together with the steady-state, capacity, production, and balanced growth constraints, form a quadratically constrained program (linear for any given $\mu$). 

To obtain the optimal resource allocation that maximizes $\mu$, a binary search over the growth rate $\mu$ is performed, and in each step a new linear program is solved, as described in \citep{ruegen2015}. Biologically speaking, if the cell can grow at rate $\mu_1$ and at a rate $\mu_2\geq\mu_1$, then it should also be able to grow at any growth rate $\mu_c$, with $\mu_1\leq\mu_c\leq\mu_2$. Thus, a binary search is an appropriate algorithm.


From a numerical perspective, the linear programs solved within the binary search are ill conditioned. Even when the constraint matrices are suitably scaled, standard commercial solvers cannot be used due to lack of numerical precision. Instead, SoPlex \citep{wunderling1996,gleixner2012}, a more stable open-source solver that can perform iterative refinement of the solution has been used. Due to the numerical instabilities, the solving routines may still take several hours up to days because of the high number of iterations necessary. Further details of the implementation are provided in the Supplementary Information.


\section*{Acknowledgements}
The work of AMR was partially funded by the International Max Planck Research School for Computational Biology and Scientific Computing in the form of a PhD stipend. RS and HK are supported by the German Federal Ministry of Education and Research as part of the “e:Bio – Innovationswettbewerb Systembiologie” [e:Bio – systems biology innovation competition] initiative (reference: FKZ 0316192). The work is part of a joint research grant funded by the Einstein Foundation Berlin (to AB and RS).


\newpage
\centerline{{\bf\Large Supplementary Material to}}
\centerline{{\Large Evaluating the stoichiometric and energetic constraints}} 
\centerline{{\Large of cyanobacterial diurnal growth}}

\setcounter{figure}{0}
\renewcommand{\thefigure}{S\arabic{figure}}

\tableofcontents
\newpage

\section{Model assumptions and limits of applicability}

The statement ``Essentially, all models are wrong, but some are useful'', attributed to the statistician George Box, also holds for our model. 
We are confident that our analysis provides a reasonable first description of the energetic
and stoichiometric constraints of phototrophic growth. 
Nonetheless, we wish to note some discrepancies and possible improvements with respect to the analysis, 
and provide additional remarks to put our results into context. 
Subsequently, we provide all necessary information relevant to implement the model. 

\begin{itemize}

\item {\bf Growth rate and multiplication factor:} Relationships between growth rate $\lambda$ (unit $h^{-1}$),
the multiplication factor $\mu$ (unitless), and the division time $T_\mathrm{D}$ (unit: $h$) are
\begin{equation}
\mu = \exp(\lambda \cdot 24\mathrm{h}), \qquad 
\lambda = \frac{\log \mu}{24\mathrm{h}}, \qquad 
T_\mathrm{D} = \frac{log(2)}{\lambda}~.
\end{equation}

\item {\bf Light intensity and the effective cross section of photosystems:} 
Within the model, light absorption and photosynthetic activity is constrained by two quantities: The effective maximal cross section of each photosystem (which depends also on attachments of phycobilisomes in the case of PSII), as well as the catalytic efficiency of each photosystem.
Absorbed light (number of photons) per photosystem corresponds to the incoming light intensity multiplied by
the effective maximal cross section. The latter values are sourced from the literature as $\sigma_\mathrm{PSII} = 1.0 \mathrm{nm}^2$ and 
$\sigma_\mathrm{PSI} = 0.5 \mathrm{nm}^2$~\cite{Mackenzie2004}.
Inspection of Figure 2A of the main text indicates that these values are too low. However, since a modification of parameters
with hindsight would violate our aim of an {\it ab initio} prediction of emergent properties,
we decided to keep Figure 2A unchanged. 
We note, however, that the absolute value of the incoming light intensity only impacts the model
via the effective cross sections. 
A change of the effective cross sections results in a shift of incoming light intensity, with no further impact
on any simulations or model-derived property. 
Since the effective cross section is quadratic as a function of diameter, small changes in the effective diameter may result
in significant changes with respect to the saturating light intensity. 
We expect a value of $1000-2000 \, \mathrm{mmol}$ photons $\mathrm{gDW^{-1} \, h^{-1}}$ to be a more realistic 
value than the current estimate, the effective diameter must then be approximately double the assumed value. 

\item {\bf Constraints and range of applicability:} The evaluation of our model is based on the assumption of
a stationary culture in a periodic environment (equation (1) within the main text). 
This equation primarily holds on the culture level. That is, under stationary conditions, we expect a
measurement of average cell composition to be invariant with respect to a full
diurnal cycle $24$h (either in a turbidostat setting or via serial dilution). 

We note, however, that equation (1) must not necessarily hold for an individual cell. 
Nonetheless, equation (1) is still a valid assumption for our analysis, based on the following arguments:
Firstly, cyanobacterial growth happens on diurnal time scales. Typical division times are approximately $24\mathrm{h}$. 
Faster rates, up to $2.5-3\mathrm{h}$ are only observed under highly optimized conditions. 
Secondly, our main interest are (metabolic) synthesis reactions related to diurnal growth. 
We conjecture that such cellular temporal programs go beyond the timespan of a single cell cycle. 
Indeed, it has been shown that the phase of cellular oscillations persists also after
division events. It is therefore reasonable to assume that an evolved 
temporal metabolic program to optimize cellular resource allocation reflects
the external (light) conditions, according to equation (1), rather than, for example, an individual cell cycle. 
Specifically, our assumption implies that a cell has evolved to synthesize glycogen 
according to the global resource allocation problem considered herein, even though dusk might happen only
after 1-2 generations (division events). 

In contrast to the scenario of fast and medium division times (of a time scale of $24\mathrm{h}$ and faster), we note
that we do not necessarily expect our analysis to capture cellular resource allocation for very slow division times. 
While equation (1) certainly remains applicable, other constraints than the energetic
implications of {\it de novo} protein synthesis (such as protein turnover and repair) become relevant,
and eventually dominant. 
Such additional constraints may be included within the model, but are outside the scope of the current analysis. 
 



\item {\bf Energy valves and cross-talk between photosystems:} 
Our model does not explicitly represent cross-talk and energy spillover between photosystems. 
We note that, despite their importance {\it in vivo}, "safety valves" of the electron transport chain are of lesser importance for our analysis. 
The analysis is based on optimal resource allocation, given a known diurnal light environment. 
Under these conditions, the system does
not have to accommodate sudden fluctuations in light (or other environmental conditions) that might require 
photoprotections and alternative electron flows. Therefore we consider it reasonable to not explicitly
represent cellular mechanisms that allow cells to deal with fluctuating conditions.
The protein investments can be considered as part of the general quota protein assignment.

\item {\bf Possible improvements of the model:} 
We consider our analysis to be a reasonable first instalment to evaluate the energetic and stoichiometric implications of diurnal phototrophic growth. Nonetheless, the model allows for a number of improvements to evaluate specific environmental
conditions in future analysis. In particular, carbon limitation might be considered which requires a more detailed representation of carbon cycling processes and the carboxysome. In its current instalment, the cost of carbon 
cycling is part of general maintenance. Likewise, limitations of other factors, in particular nitrogen, but also the availability of transition metals, may be included. We also expect that a more detailed representation of photodamage, as a result of 
high light intensity, should be considered in future instalments of the model. In each case, quantitative information about the respective processes exists. A particular challenge, however, is to formulate the respective processes such that 
a solution of the respective LP remains computationally feasible.


\end{itemize}


\section{The stoichiometric reconstruction of the cyanobacterium {\it Synechococcus elongatus} PCC 7942}

The metabolic network of {\it Synechococcus elongatus} PCC 7942 (Syc7942) was reconstructed
as outlined previously~\citep{Knoop2013} using the reconstruction of {\it Synechocystis} sp. PCC 6803 (Syn6803) as a scaffold. 
Main differences to the reconstruction of {\it Synechocystis} sp. PCC 6803 are:
(1) a smaller genome size, $3.57$ Mb for Syn6803 versus $2.8$ Mb for Syc7942;
(2) no known tocopherol synthesis;
(3) no known PHB and cyanophycin pathways;
(4) no known echinenone synthesis (carotenoid)
(5) no known delta 6 and 15 desaturases (fatty acids);
(6) no annotated urea metabolism;
(7) methionine synthesis is annotated (in contrast to Syn6803)
(8) only alternative synthesis pathway for branched chain amino acids (according to~\cite{wu2010});
as well as an incomplete TCA cycle that cannot operate in cyclic mode.
Neither the bypass of~\cite{Zhang2011}, nor the GABA shunt~\citep{Knoop2013} is annotated,
neither a malate dehydrogenase, nor a glutamate dehydrogenase is annotated. 
The reconstructed network is provided as Supplementary File 2 (SBML).

\section{The macromolecules of auto-catalytic growth}

Given the metabolic network reconstruction, we use the reaction-gene mapping together with the sequence annotations
to describe the production of each metabolic enzyme from the metabolic network. 
In the following, we describe how the production of macro-molecules was modelled, based on the
metabolic reconstruction of Syc7942.

\subsection{Metabolic Enzymes}

Each reaction in the metabolic network is either catalysed by an enzyme or is a spontaneous reaction.
For enzyme-catalysed reactions, several possibilities exist: the enzyme is encoded by only one gene,
the enzyme is a protein complex that involves several genes, there are multiple isoenzymes, 
either encoded by a single gene or as complexes, that catalyse the same reaction.
Within our model, the synthesis of enzymes is described by overall reactions that take the energy
expenditure and stoichiometries of the respective amino acids into account (that is, we do not represent transcription
and translation as individual processes). Enzyme synthesis is limited by ribosome availability (see below).

\subsubsection{Proteins and enzymes encoded by individual genes}

This is the most straightforward case. The synthesis of the respective enzyme then consists of a reaction
that consumes all the necessary amino acids, the energy and the cofactors needed
for the production of one unit of the respective enzyme. 
The information about the amino acid counts is obtained from the annotated genome sequence~\citep{Nakao2010} (\url{http://genome.microbedb.jp/cyanobase}). 
The stoichiometries for the necessary energy (ATP, GTP) and other cofactors are modelled after~\cite{Nelson2008} and
represent the energy expenditure of peptide elongation.

We take as an example the synthesis of the enzyme pyruvate kinase (EC $2.7.1.40$) which catalyses the conversion of phosphoenolpyruvate into pyruvate. In Syc7942, the enzyme is encoded by the gene Synpcc7942$\_$0098. 
Following~\citep{Nelson2008}, for each amino acid added to a growing peptide, one ATP and $2$ GTPs are needed,
resulting in one AMP, one PP$_{\mathrm{i}}$, 2 GDP and one P$_{\mathrm{i}}$.
In addition, for the translation initiation, one N$^{10}$-Formyltetrahydrofolate is required and one Met-tRNA$^{\mathrm{fMet}}$, that are converted into tetrahydrofolate and fMet-tRNA$^{\mathrm{fMet}}$.
Since we are not modeling tRNA explicitly, for our purposes only the conversion of N$^{10}$-Formyltetrahydrofolate
to tetrahydrofolate is relevant. 
The metabolites involved in the production of pyruvate kinase are shown in
Table~\ref{tbl:aa_pyruvate_kinase}. The synthesis of other enzymes is modelled accordingly. 

 \renewcommand{\thetable}{S\arabic{table}} 
 \setcounter{table}{0}
\begin{table}[h!]
\centering
    \begin{tabular}{l r}\hline
        Metabolite & Stoichiometry \\ \hline \hline
        L-Histidine & $-9$\\
        L-Phenylalanine & $-11$\\
        L-Tryptophan & $-2$\\
        L-Glutamine & $-28$\\
        L-Isoleucine & $-49$\\
        L-Glutamate & $-35$\\
        L-Aspartate & $-30$\\
        L-Threonine & $-44$\\
        L-Asparagine & $-18$\\
        L-Valine & $-64$\\
        L-Proline & $-29$\\
        L-Tyrosine & $-5$\\
        L-Serine & $-40$\\
        Glycine & $-46$\\
        L-Leucine & $-56$\\
        L-Cysteine & $-2$\\
        L-Alanine & $-60$\\
        L-Methionine & $-10$\\
        L-Arginine & $-35$\\
        L-Lysine & $-21$\\
        N$^{10}$-Formyltetrahydrofolate & $-1$\\
        ATP & $-594$\\
        GTP & $-1188$\\
        \bf{Pyruvate kinase} & $\mathbf{1}$\\
        Tetrahydrofolate & $1$\\
        AMP & $594$\\
        PP$_{\mathrm{i}}$ & $594$\\
        GDP & $1188$\\
        P$_{\mathrm{i}}$ & $1188$\\\hline
    \end{tabular}
\caption{The synthesis of the enzyme pyruvate kinase is described as a single overall process. Shown are the metabolite and energy expenditures involved in the production of one enzyme pyruvate kinase. Negative stoichiometries indicate consumption in the process, positive stoichiometries correspond to compounds that are produced. The synthesis of other enzymes is described analogously. The total rate of protein synthesis (translation) is constrained by ribosome availability.} \label{tbl:aa_pyruvate_kinase}
\end{table}

\subsubsection{Protein complexes encoded by multiple genes}\label{sec:rxnManyGenes}
For protein complexes, we need to know the stoichiometry of each protein in the complex. 
The production of the complex is otherwise identical to the case when one gene is involved. 
The total amino acid requirement is given by the sum of amino acid counts of the constituent proteins. 
Synthesis is not mechanistic, that is, the assembly of the protein complex is described 
as a single overall reaction. 

\subsubsection{Isoenzymes encoded by different genes}
For each isoenzyme a production reaction using the amino acid count of the corresponding gene is included.
In case several genes are involved (protein complexes), we proceed as in subsection~\ref{sec:rxnManyGenes}.

\subsection{Ribosomes}
Ribosome synthesis is modelled analogously to enzyme production. 
We assembled a list of the ribosomal proteins, their corresponding genes, and the ribosomal RNA, based on
the Kyoto Encyclopedia of Genes and Genomes (KEGG) resource (\url{http://www.kegg.jp/}). 
The corresponding sequences can be retrieved from databases (here: Cyanobase,\citet{Nakao2010}).
Tables~\ref{tbl:ribosome_comp1}-\ref{tbl:ribosome_comp3} provide the ribosome composition of Syc7942.


The ribosome translation rate of Syc7942 has not been measured directly, we therefore assume a
rate similar to that of \textit{Escherichia coli}, namely 12-17 amino acids per second~\citep{young1976}. 
Within the model, we use a value of $15$ amino acids/s as ribosome translation rate.
Figure~\ref{fig:ribosomeSensitivityGrowth} shows the sensitivity of modeling results with respect to the translation rate, the results are largely robust with respect to (moderate) changes in the translation rate.

\subsection{Synthesis of Photosystems and the Electron Transport Chain}

The photosynthetic electron transport chain consists of a number of protein complexes that are synthesized 
as outlined above. 
We note that the location of the protein complexes is of lesser importance for the resource 
allocation problem. Therefore, only the ETC of the thylakoid membrane is considered.
The protein complexes of the ETC constrain flux through the ETC analogously to metabolic enzymes
that constrain biochemical flux.

\subsubsection{The photosynthetic ETC}
The reactions of photosystem I (PSI) have been merged together into a single overall reaction:
$$1 \, \mathrm{photon} + 1 \, \mathrm{oxidized\;ferredoxin} + 1 \, \mathrm{reduced\;plastocyanin}$$
$$\rightarrow 1 \, \mathrm{oxidized\;plastocyanin} + 1 \, \mathrm{reduced\;ferredoxin}.$$

Similarly, the reactions of photosystem II (PSII) have been merged together into the overall reaction:
$$4 \, \mathrm{photons} + 2 \, \mathrm{H_2O}  + 2 \, \mathrm{plastoquinone} + 4 \, \mathrm{H}^+$$
$$\rightarrow
1 \, \mathrm{O}_2 + 2 ,\ \mathrm{plastohydroquinone} + 4 \, \mathrm{H}^+.$$

The reactions of the cytochrome b$_6$f (Cytb$_6$f) complex have been merged together into:
$$1 \, \mathrm{plastohydroquinone} +  2 \, \mathrm{oxidized\;plastocyanin} + 2 \, \mathrm{H}^+$$
$$\rightarrow
2 \, \mathrm{reduced\;plastocyanin} + 1 \, \mathrm{plastoquinone} + 4 \, \mathrm{H}^+.$$

The NADPH dehydrogenase complex (NDH I) is known to participate in a variety of reactions within respiration, cyclic electron transport around PSI and CO$_2$ uptake \citep{Ma2015}, its precise role is not fully understood.
In our model, NDH I catalyses the following two reactions (we note that a possible transfer of electrons via FNR from NADPH is of lesser importance for our considerations as we are mostly interested in net energy expenditure): 
$$\mathrm{NADPH}+5 \, \mathrm{H}^++\mathrm{plastoquinone}
\rightarrow
\mathrm{NADP}^++\mathrm{plastohydroquinone}+4 \, \mathrm{H}^+,$$
and
$$\mathrm{NADPH} + 4 \, \mathrm{H}^++\mathrm{plastoquinone}+\mathrm{H_2O}+\mathrm{CO}_2$$
$$\rightarrow
\mathrm{NADP}^++\mathrm{plastohydroquinone}+4 \, \mathrm{H}^++\mathrm{HCO}_3^-.$$

The gene compositions of PSI, PSII, NDH I and Cytb6f and the corresponding stoichiometries are provided in Tables~\ref{tbl:photosys_comp}-\ref{tbl:NDHCytb6f}. 
We note that pigments are necessary compounds for the synthesis of the photosystems. For pigments whose stoichiometries are not known, a separate quota metabolite is included to enforce their presence in the model biomass.

\subsubsection{Phycobilisomes}

Phycobilisomes (PBS) are protein complexes that act as light harvesting antennae and capture light of wavelength in the range of $450-650$nm~\citep{Thomas1993}. The captured energy is transferred to the PSII chlorophyll. 

In~Syc7942, phycobilisomes are essential for the correct functioning of PSII \citep{bhalerao1995}. 
PSB attach and detach from the photosystems and in this way also regulate how much light is absorbed and transferred to the photosystems~\citep{Liu2013}. 
Since PBS are assumed to be predominantly associated with PSII, we only consider energy transfer from PBS to PSII for now.

From a structural perspective, the phycobilisomes are made of two cylinders that form the core and six light harvesting antennae of variable length~\citep{bhalerao1995, campbell1998}. The antenna length influences the efficiency of the phycobilisome in the sense that the longer the antenna, the more efficient the light harvesting. In times of high light intensity, the antennae can be shortened and the proteins that belonged to them are degraded back into individual amino acids.

\begin{figure}[th]
    \centering
    \includegraphics[width=.6\textwidth]{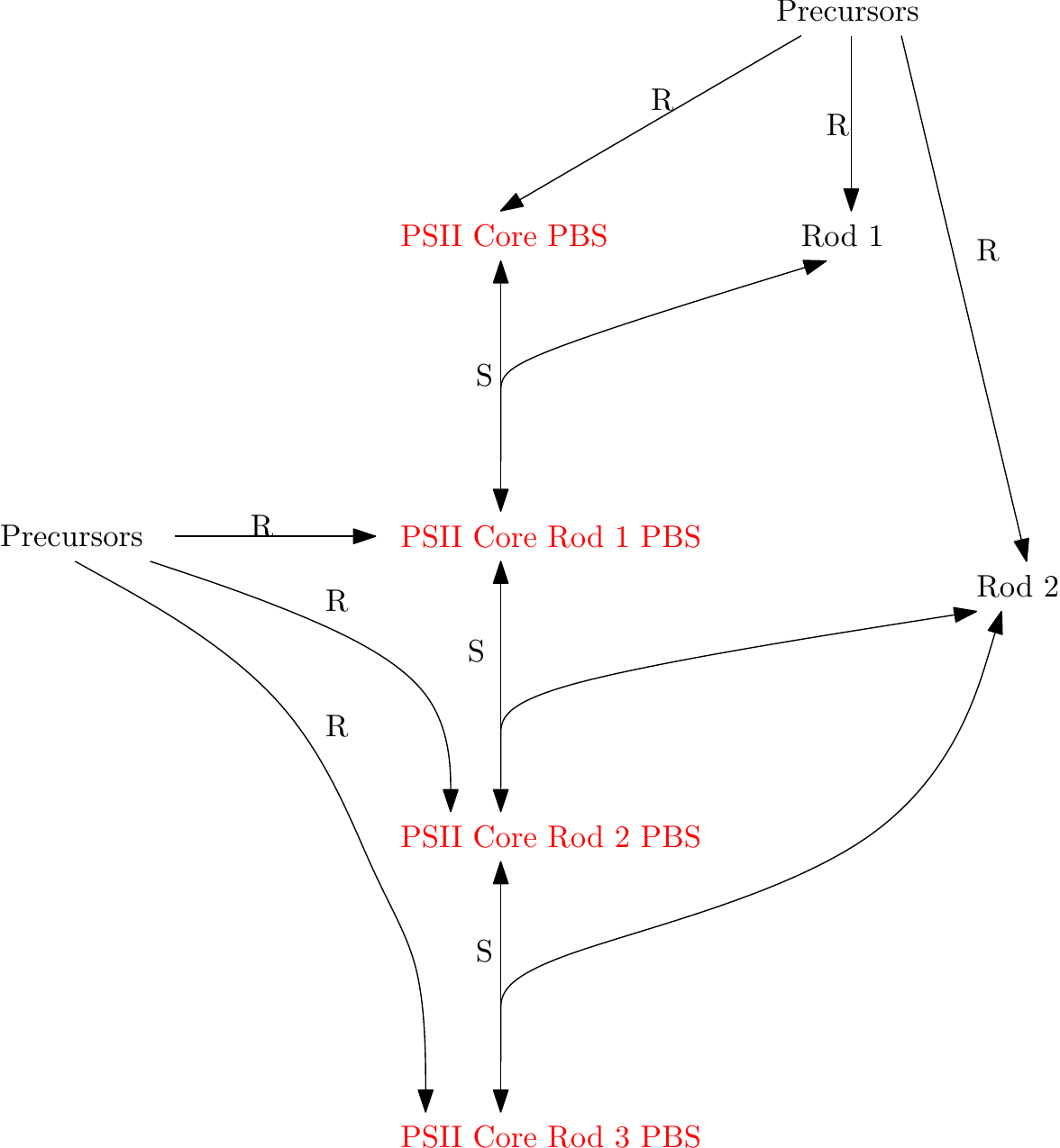}
    \caption{\textbf{PSII-phycobilisome complexes and the transitions between them.} The PSII-phycobilisome complex reactions labelled with R are assumed to be catalysed by the ribosome, while reactions labelled with S are assumed to be spontaneous. The following abbreviations are used: PSII Core PBS - Complex of photosystem II and core phycobilisome, PSII Core Rod 1 PBS - Complex of photosystem II, core phycobilisome, and 6 antennae of length 1, PSII Core Rod 2 PBS - Complex of photosystem II, core phycobilisome, and 6 antennae of length 2, PSII Core Rod 3 PBS - Complex of photosystem II, core phycobilisome, and 6 antennae of length 3, Rod 1 - rod protein made of phycocyanin, CpcA, CpcB and CpcG, Rod 2 - rod protein made of phycocyanin, CpcA, CpcB and CpcC.}\label{fig:phycobilisome_states}
\end{figure}

Based on this information, we model the photosystem II and phycobilisome in individual states, according to how long the antennae are. We consider as base state a PSII to which the core of the phycobilisome is attached.
Cyanobacteria with such phycobilisomes are able to survive at very low growth rates~\citep{bhalerao1995}. 
We then model transitions to other states where the antennae size is increased by reactions that ``consume'' the respective proteins and produce a complex with longer antennae. 
The respective transitions are shown in Figure~\ref{fig:phycobilisome_states}.
The gene composition of a phycobilisome is provided in Table~\ref{tbl:PBS_comp}.

\subsubsection{Respiratory chain}
The gene composition of ATPase, Cytochrome c oxidase, and succinate dehydrogenase are detailed in Table~\ref{tbl:resp_comp}.

\section{Model objectives and constraints}
\subsection{Notation}
Let $S$ denote the stoichiometric matrix of the system, which encompasses all metabolic and macromolecule production reactions. We denote by $S_i^j$ the stoichiometric coefficient of compound $i$ in reaction $j$.

We denote the set of internal metabolites by $\IntMet$, the set of enzymes by $\Enz$, the ribosome by $\rib$, and glycogen by $\glyc$. We call internal macromolecules with no catalytic activity (DNA, RNA, cell wall, pigments, nonmetabolic proteins, lipids, cofactors and vitamins, ions) but which are still necessary components of the biomass as~\textit{quota compounds}
and denote them by $\Quota$.
Unless otherwise specified, we consider the ribosome as an enzymatic component, i.e. $R\in\Enz$.

With respect to reactions, we distinguish between metabolic reactions $\IntRxn$, enzyme production reactions $\ERxn$, exchange (transport) reactions $\ExcRxn$, that import metabolites from or export them to the extracellular environment, and reactions $\QRxn$ that produce quota metabolites. The set of irreversible reactions in the model is denoted by $\Irr$.

We denote the vector of reaction rates (fluxes) as $\v\in\mathbb{R}^{\ExcRxn\cup\IntRxn\cup\ERxn\cup\QRxn}$ and the vector of concentrations as $\c\in\mathbb{R}^{\Enz\cup\Quota\cup\glyc}$.

In general, we index sets using subscripts. For instance $\IntMet_i$ refers to the $i$-th internal metabolite.
We use bold faced letters to denote vectors.

\subsection{Constraints}
\subsubsection{Steady state versus time-dependent quantities}
The model includes processes happening within different time scales: metabolic reactions that are fast, and macromolecule production reactions which are typically significantly slower.

Therefore, as in \citep{waldherr2015,ruegen2015}, we assume central metabolism to be at steady state, 
whereas amounts of macromolecules are dynamic time-dependent quantities.
In particular, every metabolite $\IntMet_i$ is assumed to be produced (either by transport into the cell or via metabolism) at all time points at the same rate as it is consumed (to produce either enzymes or quota metabolites). 
Therefore, we obtain the constraint
\begin{align}
\frac{d\c_\IntMet(t)}{dt}=S_{\IntMet}\v(t)=0,
\end{align}
for all internal metabolites at all time points $t$.

Enzymes and ribosomes have to be synthesized by cellular metabolism and their dynamics are governed by a
system of differential equations
\begin{align}
\frac{d\c_\Enz(t)}{dt} = S_{\Enz}\v(t) = S_{\Enz}^{\ERxn}\v_{\ERxn}(t),
\end{align}
for each enzyme or ribosome at all time points $t$. Please note that only the reactions in $\ERxn$ contribute to changes in enzyme amounts. 

Similarly, quota metabolites have to be synthesized using precursors from cellular metabolism, therefore
\begin{align}
\frac{d\c_\Quota(t)}{dt} = S_{\Quota}\v(t) = S_{\Quota}^{\QRxn}\v_{\QRxn},
\end{align}
for all quota metabolites at all time points $t$. As in the case of the enzymes, only reactions in $\QRxn$ contribute to 
quota compound synthesis.

To account for basic cell maintenance in the absence of light, we allow glycogen to accumulate and be consumed.
Therefore, glycogen amount is allowed to vary and obeys the differential equation
\begin{align}
\frac{d\c_\glyc(t)}{dt} = S_{\glyc}\v(t),
\end{align}
where $\c_G(t)$ denotes the amount of glycogen at time $t$.

\subsubsection{Enzyme amounts constrain reaction rates}\label{sec:enzRxnBounds}
Enzyme amounts constrain reaction rates within the metabolic network. 
Taking into account the Michaelis-Menten rate law
\begin{align}
v = c_e \, \kcat \cdot \frac{c_{substrate}}{K_M+c_{substrate}},
\end{align}
where $c_e$ denotes the concentration of the enzyme that catalyses the reaction, $c_{substrate}$ the concentration of the substrate, and $\kcat$ and $K_M$ are the turnover rate and the Michaelis constant respectively. 
Since
\begin{align}
\frac{c_{substrate}}{K_M+c_{substrate}} \leq 1 \, ,
\end{align}
it follows that
\begin{align}
v\leq c_e \kcat ~~.
\end{align}
Thus, given $\kcat$ and the enzyme amount $c_e$, we can use their product to give
an upper bound on the flux $v$ through the reaction. The reasoning holds analogously also for
multi-substrate rate laws. 

Turnover numbers for metabolic enzymes can be retrieved from the BRENDA database~\citep{brenda}.
A known problem in this context is the fact that recorded values for a specific enzyme spread over
several orders of magnitude. In our model we deal with this problem by using the median
value of all wild type turnover numbers reported for an enzyme.

For enzymes with no annotated values for $\kcat$, we follow the strategy used by \cite{shlomi2011}: for 
these enzymes we use the median value of all the known wild type turnover rates we found as $\kcat$. 

For irreversible enzyme-catalysed reactions, our constraint then reads
\begin{align}
\v_j(t)\leq \kcat_j\c_{\Enz_j}(t),
\end{align}
for all irreversible reactions $j$ at all time points $t$.
In the case of reversible reactions, both directions are constrained
\begin{align}
\v_j(t)\leq \kcat_j^+\c_{\Enz_j}(t),
\end{align}
where $\kcat_j^+$ is the turnover rate for the forward direction and
\begin{align}
\v_j(t)\geq -\kcat_j^-\c_{\Enz_j}(t),
\end{align}
where $\kcat_j^-$ is the turnover rate for the reverse direction. We impose these two constraints at each time point $t$ for each reversible reaction $j$. The constraints apply only for enzyme-catalysed reactions. Rates of spontaneous reactions
remain free of these bounds.

In case several reactions are catalysed by the same enzyme, their total flux weighted by the inverse of the respective turnover numbers has to be bound by the enzyme amount. Such a situation happens, for instance, in the case of the ribosome. 

Specifically, the turnover number $\kcat$ of the ribosome is different for each protein as it depends on the length of the respective protein. The turnover rate of the ribosome is about $15$ amino acids per second, which means that e.g. proteins of length $100$ amino acids, assuming they do not compete with other proteins for the ribosome, will be translated at a rate of about $540$ per hour.


The ribosome, however, has to translate all proteins. Therefore, we obtain the constraint
\begin{align}
\sum_{j\in\ERxn}\frac{\v_j(t)}{\kcat_j}\leq \rib(t),
\end{align}
at each time point $t$. Note that enzyme production reactions are irreversible and hence the bound is only applied for the forward direction of the reaction, while the reverse direction is prohibited using a lower flux bound of $0$ applied at all time points.
A similar constraint applies to reactions that are catalysed by the same enzyme.

\subsubsection{Amounts of quota compounds}

The model includes $8$ quota compounds (DNA, RNA, cell wall, pigments, nonmetabolic proteins, lipids, cofactors and vitamins, ions) that are modelled dynamically. Their initial values are set to be equal to their corresponding amounts in $1$ gram dry weight of Syc7942 cells and displayed in Table~\ref{tbl:initial_quota}. We denote their initial amounts as the vector $\mathbf{q}_0$, and therefore we have the constraint
\begin{align}
\c_\Quota(t_0)= \mathbf{q}_0.
\end{align}

We additionally impose that, together with metabolic proteins and glycogen, the initial composition vector adds up
to one gram in terms of weight, and thus obtain the additional constraint
\begin{align}
\sum_{i\in\Quota}\c_i(t_0)+\sum_{i\in\Enz}\c_i(t_0)\cdot MW_i + \c_\glyc(t_0)= 1.\label{eq:sum_up_to_1}
\end{align}
We note that the protein amounts are expressed in mmol/gDW, while the glycogen and the quota amounts are expressed in grams. This is the reason why in the constraint above we need to multiply the enzyme amounts with their corresponding molecular weights $MW_i$.

\subsubsection{Adjustment of pigment quota}
Chlorophyll, $\beta$-carotene and phylloquinone are important ingredients of the photosystems. Thus, the reactions building the photosystem also incorporate these pigments. To account for this, the general pigment quota requirements have to be adjusted. Without the adjustment, the original biomass requirements are~\citep{Knoop2013}

$$0.841\; mmol\; \mathrm{Chlorophyll\; a} + 0.136\; mmol\; \beta\mathrm{-Carotene}$$
$$+ 0.321\; mmol\; \mathrm{Zeaxanthin} + 0.064\; mmol\; \gamma\mathrm{-Carotene}$$
$$+ 0.068\; mmol\; \mathrm{Phylloquinone} \to 1\; g\; \mathrm{Pigment}, $$

and the initial pigment quota would be $0.0244\; g/gDW$.

Since chlorophyll a, $\mu$-carotene and phylloquinone are no longer part of the quota compounds,
we identify the factor $f$ such that
$$f \cdot 0.321\; mmol\cdot MW(\mathrm{Zeaxanthin})+ f \cdot 0.064\; mmol\cdot MW(\gamma\mathrm{-Carotene}) = 1\;g\;\mathrm{Pigment}$$
and then change the initial pigment quota to $\frac{0.0244}{f}\; g/gDW$.
We obtain $f = 4.609$, and therefore the pigment quota formation equation becomes
$$1.479\; mmol\; \mathrm{Zeaxanthin} + 0.295\; mmol\; \gamma\mathrm{-Carotene}\to 1\;g\;\mathrm{Pigment},$$
and the initial pigment quota is $0.0053$.

\subsubsection{Periodicity of the system}\label{sec:periodicity}
We consider diurnal growth as a periodic system.
Hence, we have to enforce that for dynamically modelled variables, i.e., glycogen, ribosome, enzymes and quota metabolites, 
the amounts at the end of the time period are multiples of their amounts at the beginning of the time period.
We thus obtain the constraints
\begin{align}
\mu\c_\Enz(t_0)=\c_\Enz(t_{end}),
\end{align}
and
\begin{align}
\mu\c_\Quota(t_0)=\c_\Quota(t_{end}).
\end{align}

These constraints ensure balanced growth of the whole system as already described by \cite{ruegen2015}, see
also remarks below.

\subsubsection{Non-catalytic proteins and constraints on initial protein amounts}

The dynamics of all catalytic compounds are modelled explicitly using time-dependent balance equations. 
However, cells not only produce proteins with catalytic activity, but also other non-catalytic (quota) proteins.
To account for proteins without catalytic activity within our model, we include additional proteins 
as a quota compound, the synthesis of which is also catalysed by the ribosome. 

According to previous quantitative proteomics data by~\cite{guerreiro2014}, the proteins included within our model, 
together with the ribosomes, make up a fraction of $45\%$ of the total proteome of Syc7942.
In the original biomass reaction of the metabolic model, proteins make up $0.51$g in a gram dry weight of cells.
Of these, $45\%$ are catalytic proteins, and the remaining $55\%$ represent quota proteins. 

The model is able to choose, as part of the optimization, the initial distribution of catalytic proteins. 
To ensure that the total protein components, together with the other biomass components, sums
to one gram dry weight (gDW), equation \eqref{eq:sum_up_to_1} applies. 
For fast growing cells of Syc7942 no experimental estimates are available.
We conjecture that, based on the results, the quota of non-catalytic proteins for fast growing Syc7942 may
be significantly lower than $55\%$. Growth rate increases significantly with a decreasing amount of
quota compounds (See Figure~\ref{fig:GrowthQuota} and discussion in the main text). 
We note that minimal models of (heterotrophic) cellular growth typically also include a growth-independent
fraction of protein, typically of the order of $50\%$~\citep{Scott2014}.

\subsubsection{Light uptake}
Light availability is modelled by a (half-wave rectified) sine function that mimics the day-night cycle
\begin{align}
    l(t_i) = \begin{cases}
                 l_{max}\sin\left(\frac{2\pi t_i}{T_\mathrm{day}}\right)&\mbox{if } \sin\left(\frac{2\pi t_i}{T_\mathrm{day}}\right)\geq 0\\
                 0&\mbox{else},
             \end{cases}
\end{align}
where T$_\mathrm{day} = 24$h and $l_{max}$ the maximum light intensity that occurs T$_\mathrm{day}/4$ hours after dawn.

The amount of light absorbed by the system is proportional to the combined amount of photosystem I and photosystem II-phycobilisome complexes at the respective time point multiplied by their respective effective cross sections:
\begin{align}
    \v_{P700}(t) &\leq \sigma_\mathrm{PSI}\cdot \c_{PSI}(t)\\
    \v_{P680}(t) &\leq \sigma_\mathrm{PSII}\cdot \c_{PSII}(t),
\end{align}
where $\v_{P700}$ and $\v_{P680}$ are the fluxes of absorbed photons respectively, and $\sigma_\mathrm{PSI}$ and $\sigma_\mathrm{PSII}$ are the effective respective cross sections of photosystems-phycobilisome complexes. We do not distinguish between photons of
different wavelength (albeit it is possible to include wavelength-dependent effective cross sections, but such a consideration is beyond the scope of this work).  

The cross section of PSI is assumed to be equal to $0.5$ nm$^2$, independent of the status of phycobilisomes. 
The effective cross section of PSII depends on the length of the rods of its attached phycobilisome.
A PSII with only the core of the phycobilisome is assumed to have a cross section of $0.1$ nm$^2$, if the rods of the phycobilisome have length one, then the assumed cross section is $0.33$ nm$^2$, rods of length two give a cross section of $0.67$ nm$^2$, and full length rods give a cross section for PSII of $1$ nm$^2$~\cite{Mackenzie2004}.
The effective cross sections only affect absorbed light versus incoming light intensity and do not qualitatively 
affect simulation results.

\subsubsection{Maintenance}
Similar to conventional FBA models, we assume that there are other processes that require energy and are not considered by our model. Therefore, the model contains a non-growth associated maintenance reaction that hydrolyses ATP into ADP and P$_{\mathrm{i}}$. We enforce a lower bound of $0.13$ mmol$\cdot$gDW$^{-1}\cdot$ h$^{-1}$ for the flux through this reaction at each time point in order to account for energy consumption of general maintenance.

\subsubsection{Discretization of time points across the diurnal cycle}

To be able to formulate the optimization problem as a linear program, time is discretized. For this purpose, we use a Gau\ss{} implicit method, namely the midpoint rule. 
For a detailed explanation of this discretization rule we refer the reader to~\cite{deuflhard2002}. 
Since the ordinary differential equations are stiff, we used an implicit method for discretization.
Also other discretization methods such as the one described in~\citep{waldherr2015} can be used. 
Our choice is motivated by the size of the resulting linear program (tens of thousands of variables and constraints), as
well as by the condition of our problem.

For the $24$-hour interval we use $N=24$ discretization points. We therefore discretize, for instance, the 
time-dependence of the amount of compound $j$ as
\begin{align}
\c_j(t_i)\simeq\c_j(t_{i-1})+d\cdot\dot\c_j\left(\frac{t_i+t_{i-1}}{2}\right),\hspace{1cm} i\in\{1,\dots,N\},
\end{align}
where $d=24/N$

For the first time step we then have a special case, since $\c_j(t_0)$ is also a variable in our model.

The derivatives and the fluxes are evaluated in the middle of each time interval $[t_{i-1},t_i]$ and for the bounds described in Section~\ref{sec:enzRxnBounds} we have the set of inequalities
\begin{align}
\v_j\left(\frac{t_i+t_{i-1}}{2}\right)\leq \kcat_j\c_j\left(\frac{t_i+t_{i-1}}{2}\right).
\end{align}
Since enzyme concentrations at time points $\frac{t_i+t_{i-1}}{2}$ are not available, we approximate the previous constraint as
\begin{align}
\v_j\left(\frac{t_i+t_{i-1}}{2}\right)\lesssim \kcat_j\frac{\c_j(t_i)+\c_j(t_{i-1})}{2},
\end{align}
for each enzyme-catalysed reaction $j$ and $i\in\{1,\ldots,N\}$.

\subsection{The optimization objective}

As already introduced by~\cite{ruegen2015}, we consider a system growing in a stationary day-night culture (for example using
regular dilution after each 24h cycle). We therefore assume that the system has a period of $24$ hours, in the sense that the composition of the system at time $t_{end} = t_0+24h$ is a multiple of the composition of the system at time $t_0$. 
The assumption corresponds to balanced growth of all cellular components over a full diurnal cycle (see below for further
discussion of the periodicity constraint).

As optimization objective, we assume that the cell has evolved to grow as much as possible within a full diurnal period, that is, the regulatory system has evolved such that the factor $\mu$ involved in the constraints in Section~\ref{sec:periodicity} (equation (1) in the main text) is maximal.
We note that even if this assumption of optimality turns out to be incorrect, the optimal solution with respect to the
assumption is still of high interest to compare with experimentally observed behaviour. 
It is only by knowledge of optimal solutions of the resource allocation problem that suboptimal behaviour, or 
incorrect assumptions and parameters, can be identified. 

We start our simulation with $1$ gram dry weight and track the changes in the cellular composition and growth over day-night cycles.

\subsection{The linear program (LP) and binary search}

The optimization problem is given by
\begin{align*}
\max_{\mu,\v,\c}\;&\mu&\\\notag
\mbox{s.t.}\;&S_{\IntMet}\v\left(\frac{t_j+t_{j-1}}{2}\right)=0,&\forall j\in\{1,\ldots,N\},\\\notag
&\dot\c_\Enz\left(\frac{t_j+t_{j-1}}{2}\right) = S_{\Enz}\v\left(\frac{t_j+t_{j-1}}{2}\right),\;&\forall j\in\{1,\ldots,N\},\\\notag
&\dot\c_\Quota\left(\frac{t_j+t_{j-1}}{2}\right) = S_{\Quota}\v\left(\frac{t_j+t_{j-1}}{2}\right),\;&\forall j\in\{1,\ldots,N\},\\\notag
&\dot\c_\glyc\left(\frac{t_j+t_{j-1}}{2}\right) = S_{\glyc}\v\left(\frac{t_j+t_{j-1}}{2}\right),\;&\forall j\in\{1,\ldots,N\},\\\notag
&\sum_{i\in\mathcal{V}_{j}}\frac{\v_i\left(\frac{t_k+t_{k-1}}{2}\right)}{\kcat_i^+}\leq \frac{\c_j(t_k)+\c_j(t_{k-1})}{2},\;&\forall k\in\{1,\ldots,N\},\;\forall{j\in\Enz},\\\notag
&-\sum_{i\in\mathcal{V}_{j}\setminus\Irr}\frac{\v_i\left(\frac{t_k+t_{k-1}}{2}\right)}{\kcat_i^-}\leq \frac{\c_j(t_k)+\c_j(t_{k-1})}{2},\;&\forall k\in\{1,\ldots,N\},\;\forall{j\in\Enz},\\\notag
&\c_\Quota(t_0) = q_0,&\\\notag
&\mu\c_\Enz(t_0)=\c_\Enz(t_N),\;&\\\notag
&\mu\c_\glyc(t_0)=\c_\glyc(t_N),\;&\\\notag
&\mu\c_\Quota(t_0)=\c_\Quota(t_N),\;&\\\notag
&\sum_{i\in\Quota}\c_i(t_0)+\sum_{i\in\Enz}\c_i(t_0)\cdot MW_i + \c_\glyc(t_0)= 1,&\\\notag
&\v_{light}\left(\frac{t_j+t_{j-1}}{2}\right)\leq l\left(\frac{t_j+t_{j-1}}{2}\right),\;&\forall j\in\{1,\ldots,N\},\\\notag
&\v_{P700}\left(\frac{t_j+t_{j-1}}{2}\right) \leq A_{PSI}\cdot \c_{PSI}(t),\;&\forall j\in\{1,\ldots,N\},\\\notag
&\v_{P680}\left(\frac{t_j+t_{j-1}}{2}\right) \leq A_{PSII}\cdot \c_{PSII}(t),\;&\forall j\in\{1,\ldots,N\},\\\notag
&\v_{maintenance}\left(\frac{t_j+t_{j-1}}{2}\right)\geq 0.13,\;&\forall j\in\{1,\ldots,N\},\\\notag
&\v_{\Irr}\left(\frac{t_j+t_{j-1}}{2}\right)\geq 0,\;&\forall j\in\{1,\ldots,N\},\\\notag
&\c(t_i)=\c(t_{i-1})+d\cdot\dot\c\left(\frac{t_i+t_{i-1}}{2}\right),\;&\forall i\in\{1,\dots,N\},\\\notag
&\c_j(t_i)\geq 0&\forall i\in\{1,\ldots,N\},\; \forall j\in G\cup\Enz.
\end{align*}
where $\mathcal{V}_{j}$ denotes the set of reactions that are catalysed by enzyme $j$, and $MW_i$ the molecular weight of enzyme $i$.

We notice that, because $\mu$ and $\c$ are both variables in the model, our program contains two quadratic constraints, namely the ones that guarantee the periodicity of the system. 
Since we aim to maximize $\mu$, we run a binary search and for each new value of $\mu$ we test the feasibility of the resulting linear program as in~\citep{ruegen2015,obrien2013,goelzer2011}.


We note that the LP involves several orders of magnitude differences in the coefficients of the
constraint matrix (some variables exhibit rapid change, whereas others change slowly). 
For numerical stability, we therefore perform a scaling of some of the coefficients, which
is explained in detail by~\cite{waldherr2015}.

Due to the numerical condition of our problem, we chose to solve the individual feasibility linear programs using the SoPlex $2.2.1$ optimization package~\citep{wunderling1996, gleixner2012} (\url{http://soplex.zib.de/}), which allows iterative refinement of the solution and very precise feasibility and optimality tolerances.

All code is provided at \url{https://sourceforge.net/projects/cfba-synpcc7942/}

\clearpage

\section{SUPPLEMENTARY FIGURES}

\subsection{Supplementary Figure S2}
\vspace{3cm} 
\begin{figure}[h!]
    \centering
    \includegraphics[width=.45\textwidth]{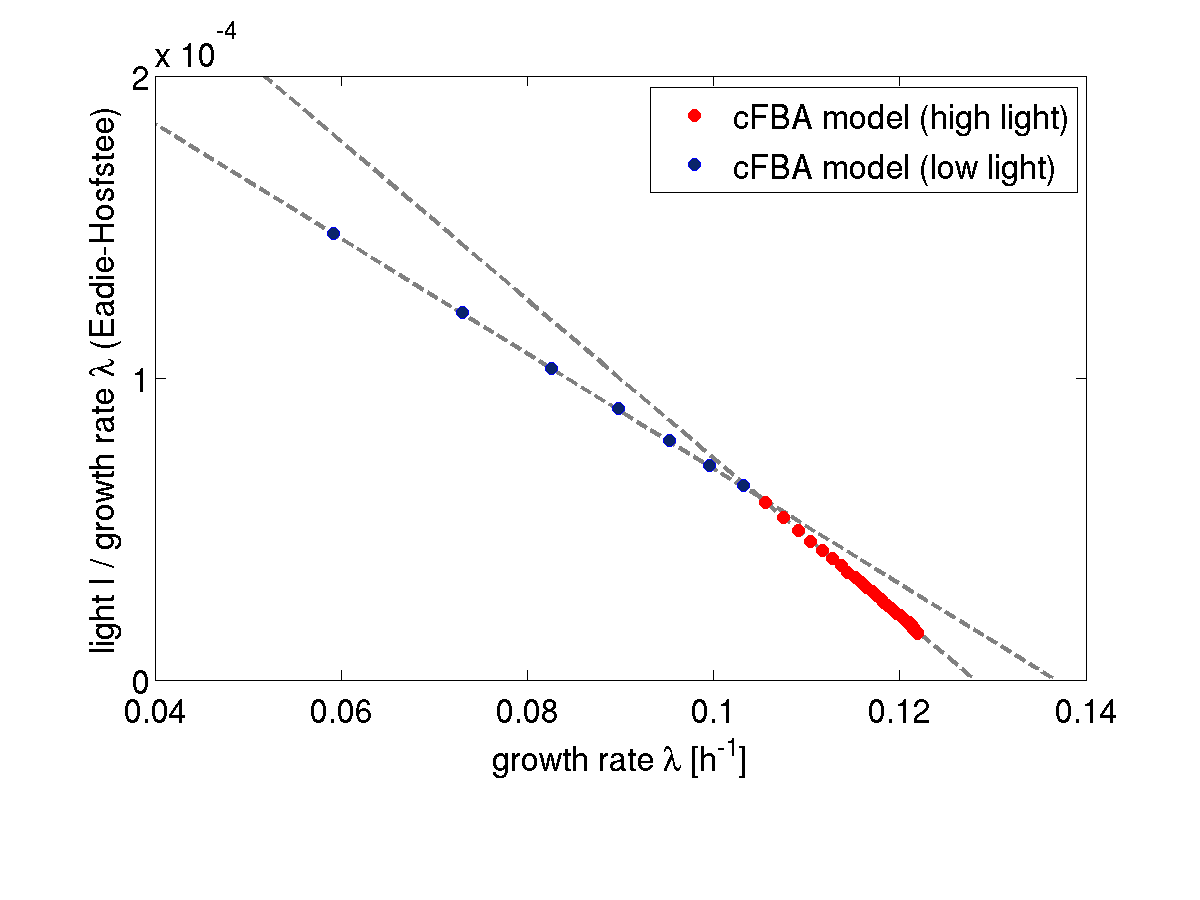}~~~
    \includegraphics[width=.45\textwidth]{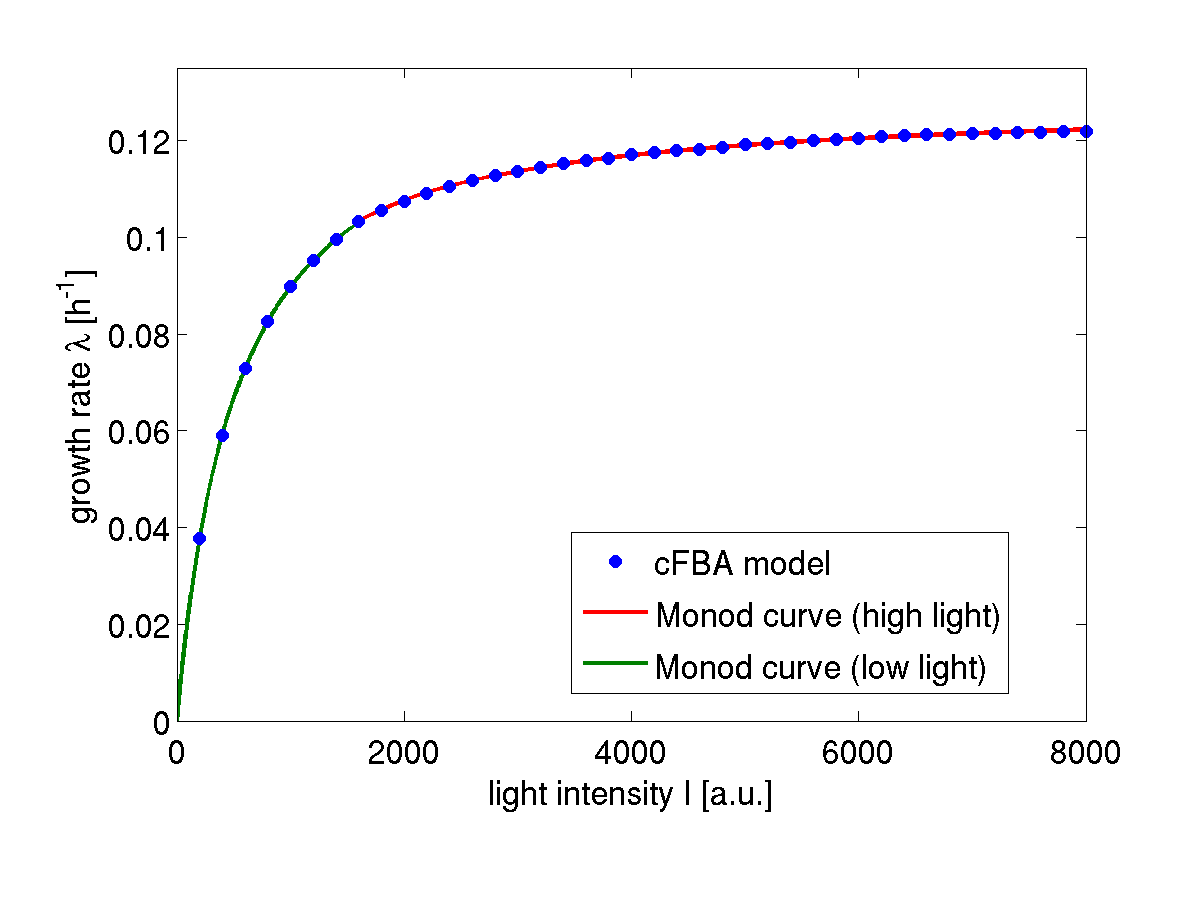}
    \caption{
{\bf The growth rate $\lambda$ as a function of light is consistent with a (bi-phasic) Monod growth equation.}
Left plot: Eadie-Hofstee plot. 
Monod growth of the form $\lambda = \frac{\lambda_\mathrm{max} \cdot I}{K_\mathrm{M} + I}$ gives rise to 
a linear relationship between $I/\lambda$ and $\lambda$ (Eadie-Hofstee plot).  
In the model, we observe bi-phasic growth and a transition at a light intensity $I \approx  1600\;\mu$mol photons $\cdot$ s$^{-1}$ $\cdot$ m$^{-2}$.
Right plot: The resulting Monod growth curve $\lambda$ versus light intensity $I$ using a linear
fit in both growth regimes (MATLAB R2016a function {\tt polyfit}), the resulting values
are $K_\mathrm{M} = 525.8$ and $\lambda^\mathrm{max} = 0.1368h^{-1}$ (light $I<1600$), as well as
$K_\mathrm{M} = 381.8$ and $\lambda^\mathrm{max} = 0.1281h^{-1}$ (light $I>1600$). We note that the absolute value
of the light intensity is solely determined by the assumed effective cross sections. 
The transition between the growth regimes is due to the transition in the limits of light capture: For low light, the activity of the photosystems is limited by the incoming light intensity, multiplied by the effective cross section. 
For high light intensity, the maximal catalytic turnover rate of each photosystem is limiting (Table 1 in main text), the absorption itself is saturated. A similar bi-phasic behavior is observed for the cellular composition (Supplementary Figure~\ref{fig:CellComp}). 
\label{fig:MonodCurve}}
\end{figure}

\clearpage 
\subsection{Supplementary Figure S3}
\vspace{3cm} 

\begin{figure}[h!]
    \centering
    \includegraphics[width=.42\textwidth]{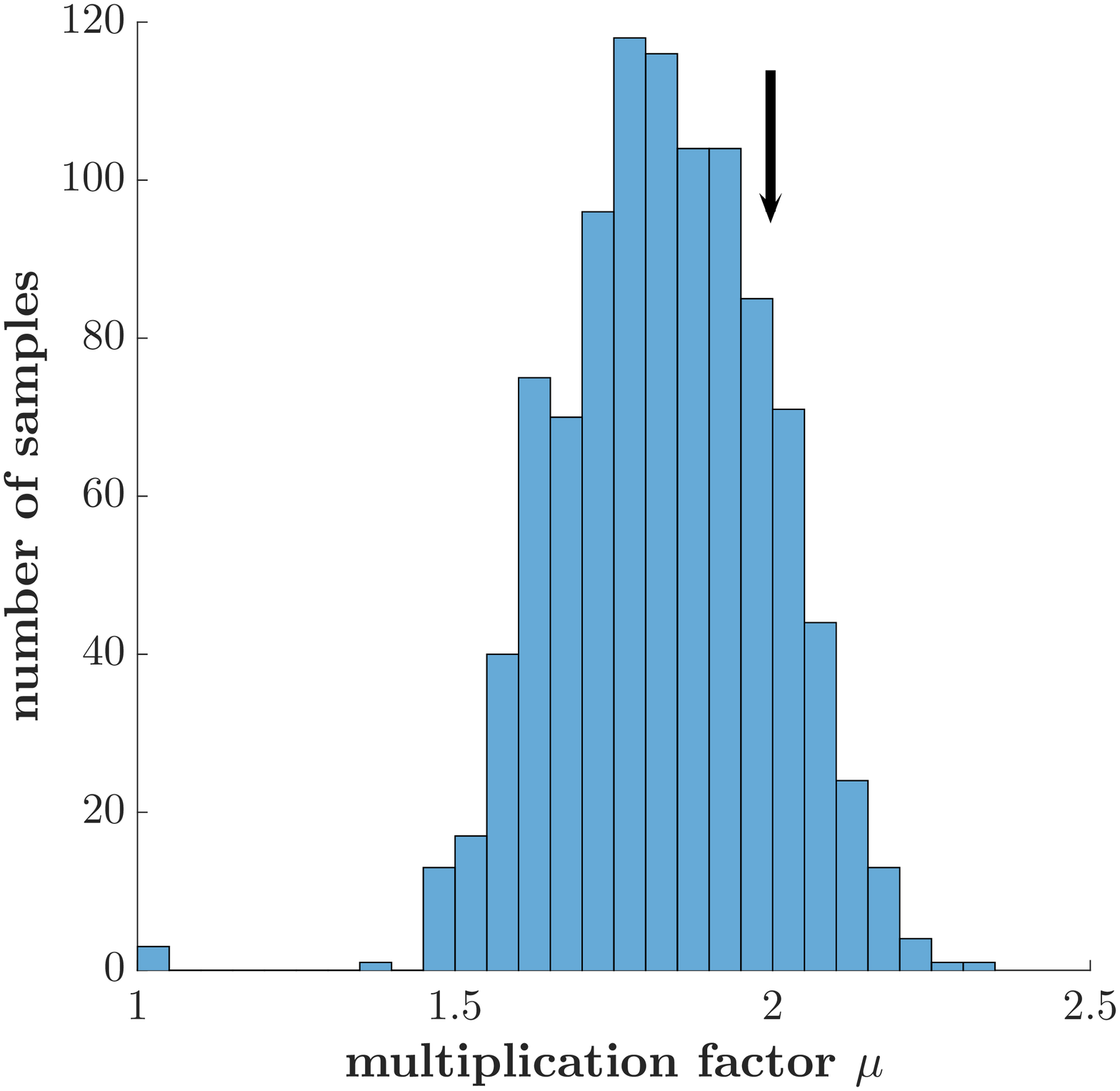}~~~
    \includegraphics[width=.48\textwidth]{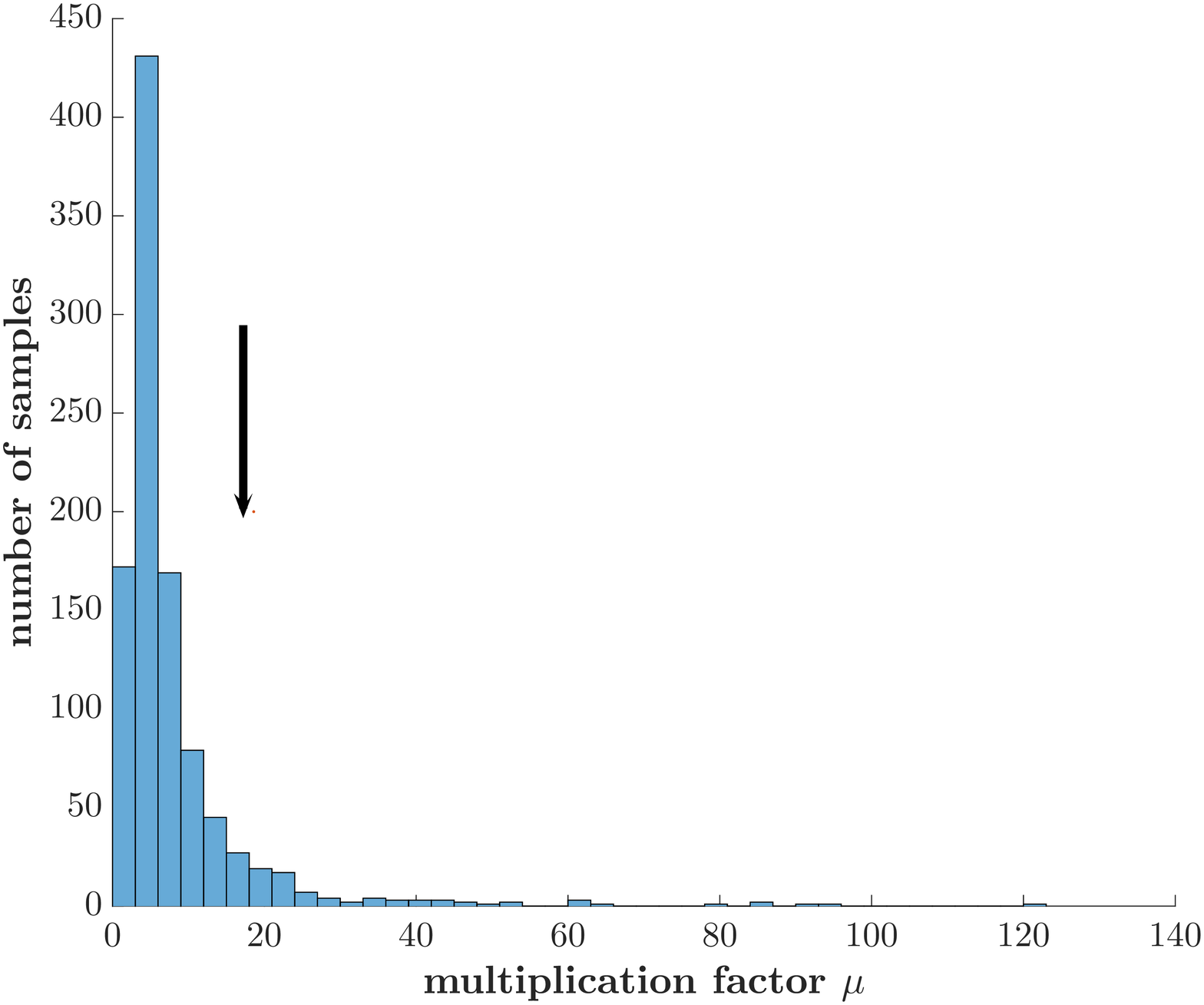}
    \caption{
{\bf Sensitivity of the growth rate to assigned turnover numbers $\kcat$ for constant light (left plot: low light, right: high light).}
To test for the importance of individual $\kcat$, we randomized the assignment between the $\kcat$ and their
respective enzymes. That is, enzymes are assigned with a $\kcat$ drawn from the original
distribution (random sampling with replacement). The randomization is motivated by the assumption that there is no
systematic bias in BRENDA as far as the overall distribution of $\kcat$ is concerned, but individual assignments may be erroneous. 
The figure shows the distribution of the multiplication factor $\mu$ after randomization.
For a light intensity of $150\;\mu$mol photons $\cdot$ s$^{-1}$ $\cdot$ m$^{-2}$ (low light, left plot) we observe 
a low sensitivity. 
The median $\mu_\mathrm{median} = 1.83$ of the distribution is close to
the original value of $\mu \approx 1.99$ (before randomization), which is indicated by the arrow. 
The fraction of randomized $\mu$ larger than the reference value is $18.2$\% (no significant difference between original and randomized
multiplication factors).
For a light intensity of $6000\;\mu$mol photons $\cdot$ s$^{-1}$ $\cdot$ m$^{-2}$ (high light, right plot) we
observe a highly skewed distribution.
Notably, the median of the distribution $\mu_\mathrm{median} = 5.0$ is markedly lower than 
the original reference value of $\mu \approx 18.6$ ($\lambda \approx 0.12$), which is indicated by the arrow.
The fraction of randomized $\mu$ larger than the reference value is $7.2$\%.  
This difference does not allow claiming a significant difference between original and randomized growth rates. 
Nonetheless, models with randomly assigned $\kcat$ seem to have consistently lower growth
rates than the original model. This fact is more pronounced at high light intensities. 
We may therefore hypothesize that the assignments of $\kcat$ are not random, but (evolutionarily) selected
to allow for higher growth rates. The hypothesis requires further investigation. 
With respect to overall sensitivity, we conclude that, at low light intensities, randomization of $\kcat$ has
no major effect on growth rate and our results are robust. 
The maximal growth rate, however, exhibits higher variability. Nonetheless, the overall variability is lower than
for changes in quota components (see Figure~\ref{fig:GrowthQuota}). 
\label{fig:ConstantLight}}
\end{figure}


\subsection{Supplementary Figure S4}
\vspace{3cm} 
\begin{figure}[h!]
    \centering
     \includegraphics[width=.6\textwidth]{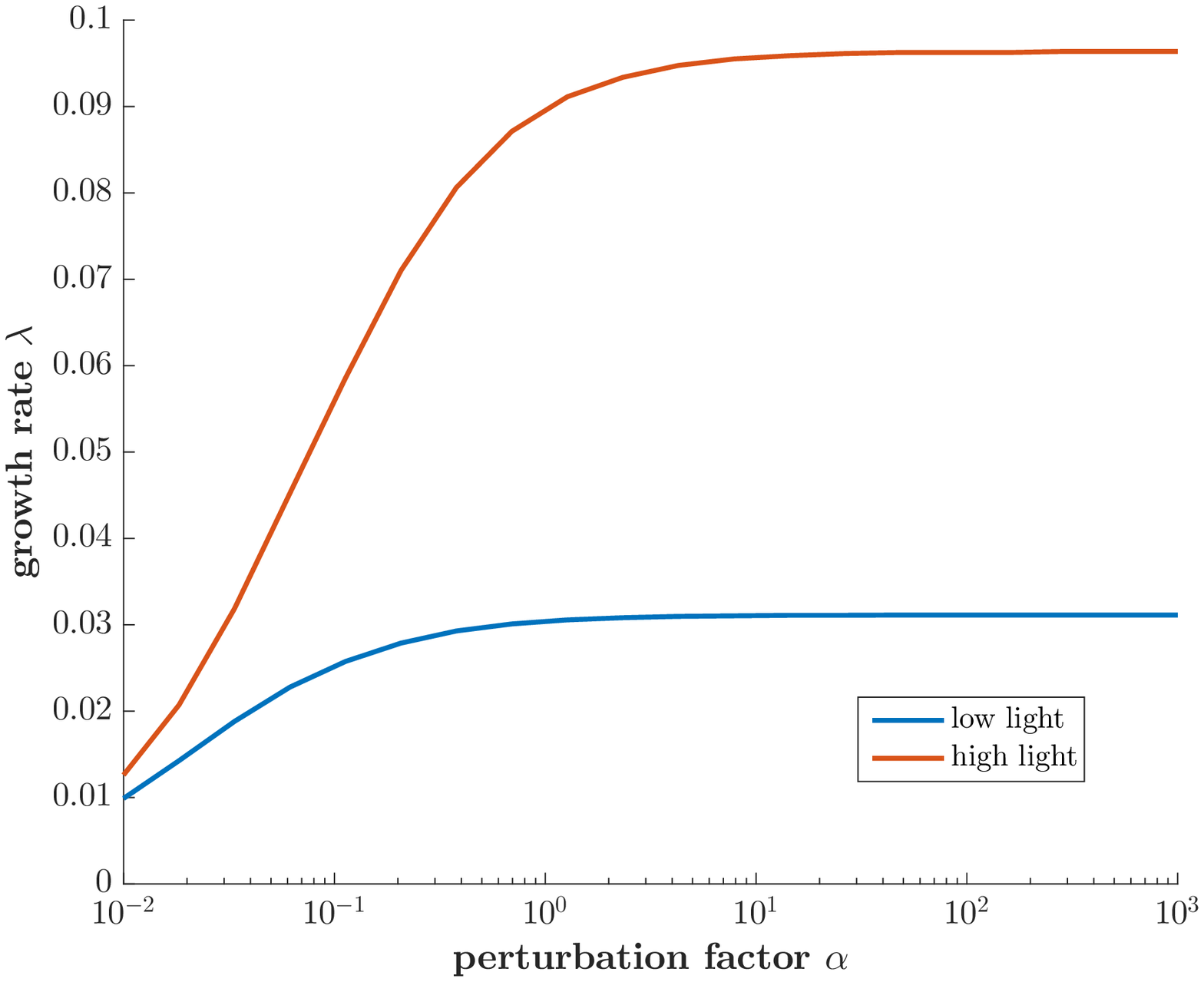}
\caption{
{\bf Dependency of the growth rate $\lambda$ on RuBisCO in constant light.}
Shown is the sensitivity of growth rate $\lambda$ to changes in the turnover rate of RuBisCO in a constant light environment. The turnover rate $\kcat^\mathrm{mod} = \alpha \cdot \kcat^\mathrm{orig}$ in the model was multiplied with a perturbation factor $\alpha$ spanning several orders of magnitude. 
We note that the growth rate does not increase significantly even for $\alpha \simeq 100$.
The simulation was run for a constant light intensity of $150\;\mu$mol photons $\cdot$ s$^{-1}$ $\cdot$ m$^{-2}$ (low light) and of $1000\;\mu$mol photons $\cdot$ s$^{-1}$ $\cdot$ m$^{-2}$ (high light).}
\end{figure}

\clearpage
\subsection{Supplementary Figure S5}
\vspace{3cm} 
\begin{figure}[h!]
    \centering
    \includegraphics[width=.5\textwidth]{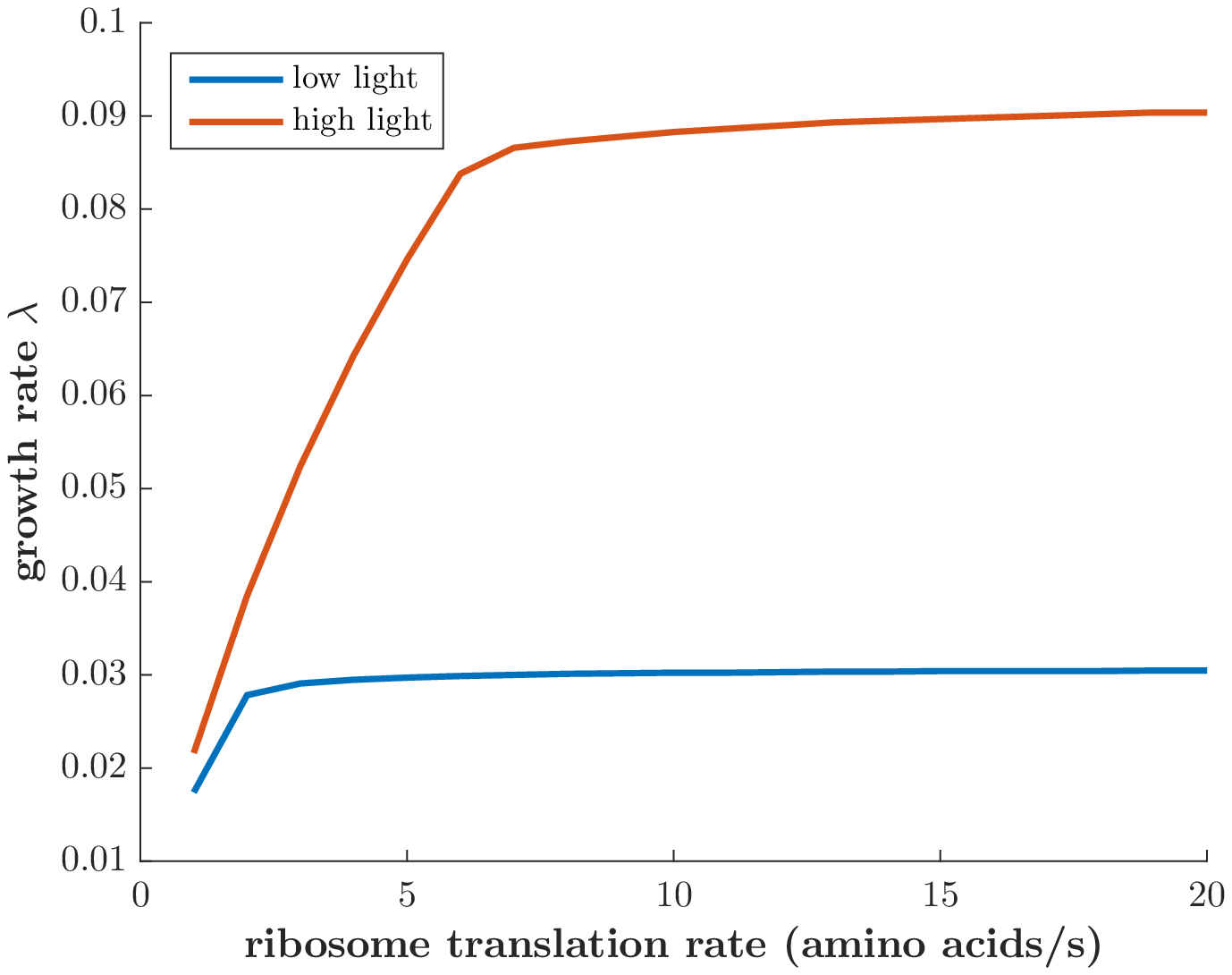}
    \caption{
{\bf Dependency of the growth rate $\lambda$ on the ribosome translation rate in constant light.}
The reference value for the ribosome translation rate used in the simulations is $15$ amino acids per second.
The simulation was run for a constant light intensity of $150\;\mu$mol photons $\cdot$ s$^{-1}$ $\cdot$ m$^{-2}$ (low light) and of $1000\;\mu$mol photons $\cdot$ s$^{-1}$ $\cdot$ m$^{-2}$ (high light).}
\label{fig:ribosomeSensitivityGrowth}
\end{figure}

\clearpage
\subsection{Supplementary Figure S6}
\vspace{3cm} 
\begin{figure}[h!]
    \centering
    \includegraphics[width=.45\textwidth]{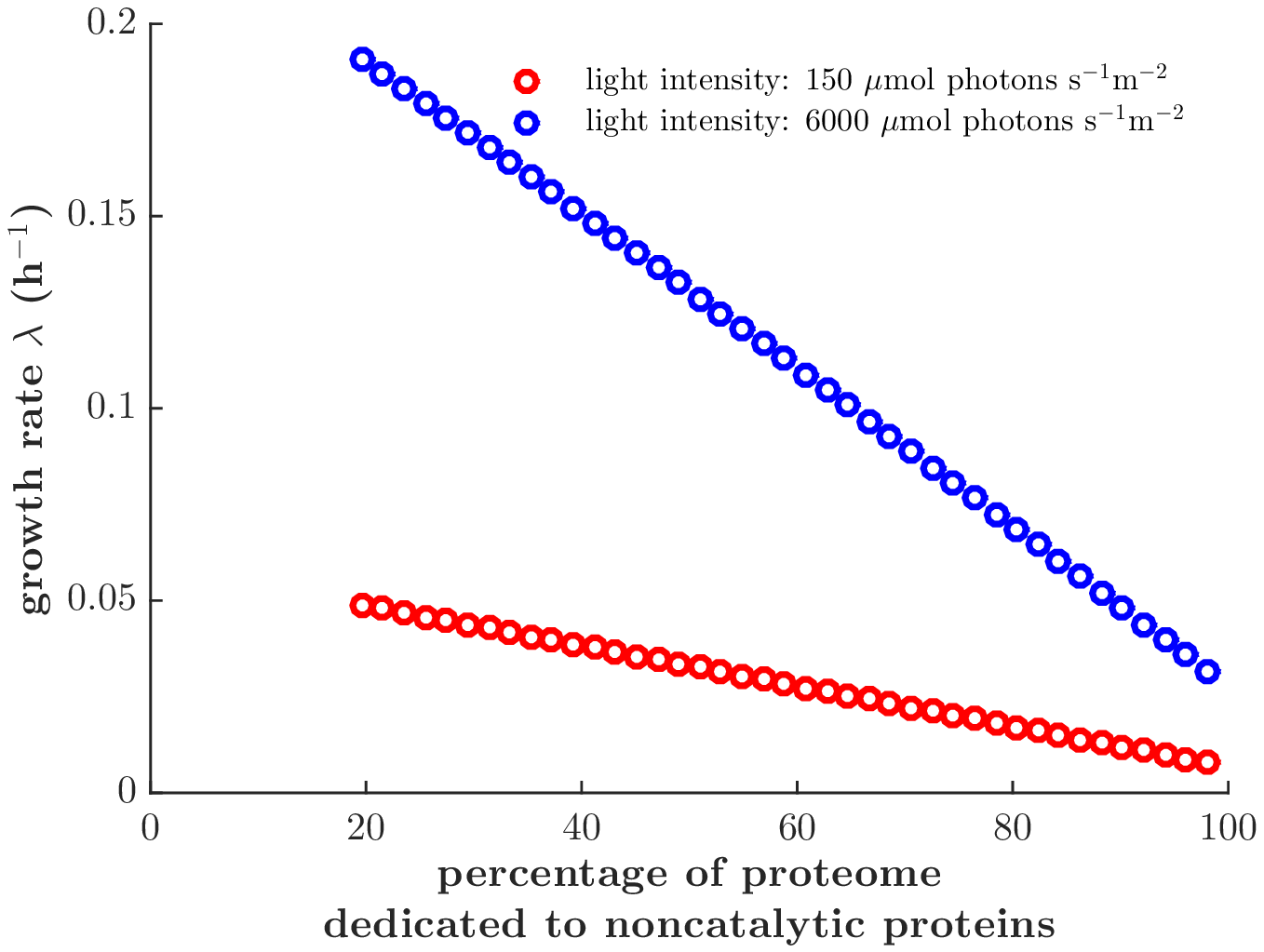}
    \includegraphics[width=.45\textwidth]{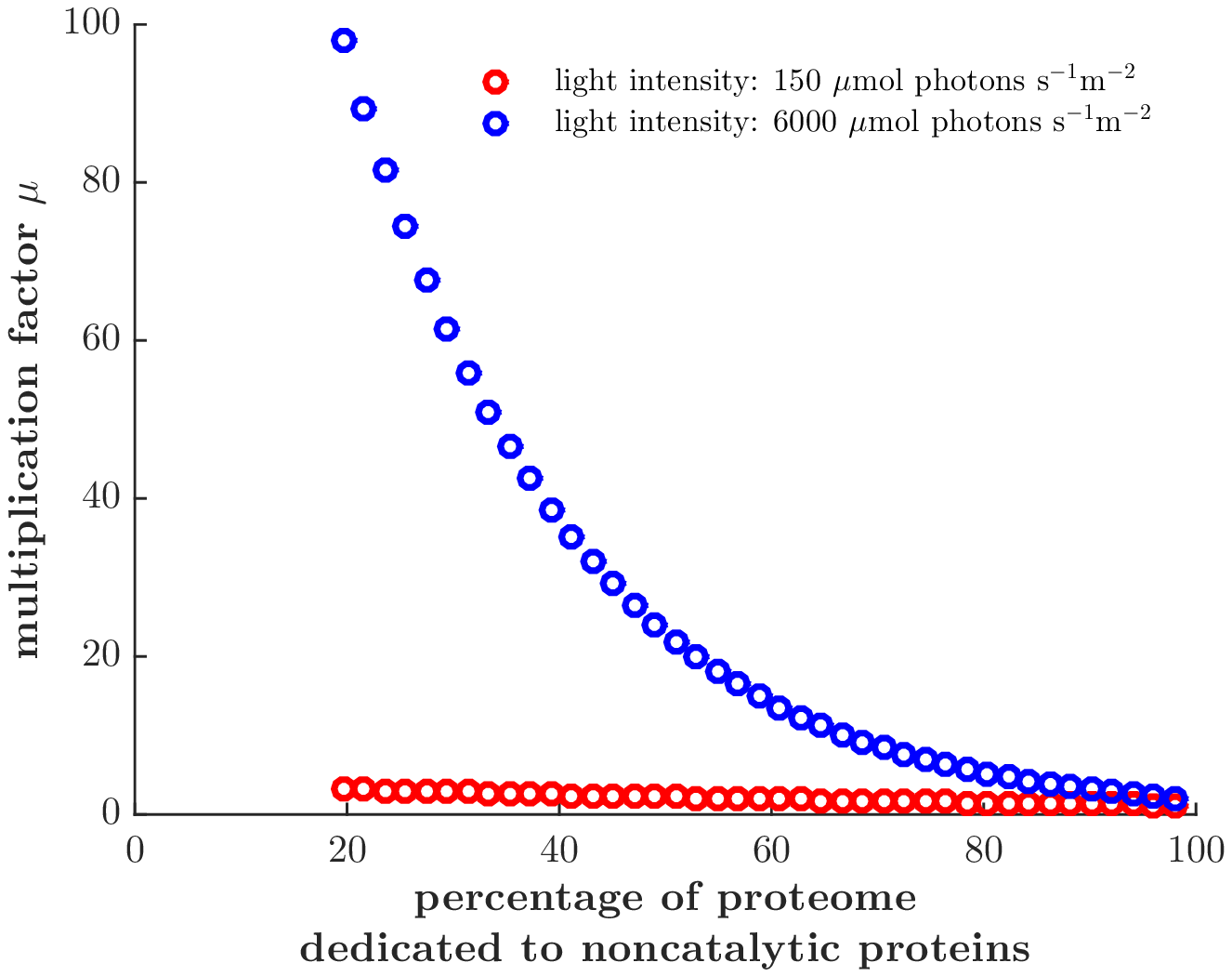}
    \caption{
{\bf Sensitivity of the growth rate on the ratio to non-catalytic proteins.}
Shown is the growth rate $\lambda$ (left plot) and the multiplication factor $\mu$ (right plot) and as a
function of non-catalytic proteins (quota compounds, i.e., proteins that serve no
catalytic function within the model and are part of the fixed protein quota). 
As expected, the growth rate increases for a lower amount of quota proteins. 
Recent proteomics data suggests a percentage of $55$\% of total protein~\citep{guerreiro2014}, measured for
slow growing cells. The value $55$\% was used as a reference value in all simulations. 
We conjecture, however, that the amount is variable and considerably lower for fast growing cells. 
We note that the protein complexes of the ETC, phycobilisomes, and proteins of central metabolism (including RuBisCo) are
assumed to constitute the bulk of the proteome and are all included as catalytic proteins within our model. 
If the quota of non-catalytic proteins is assumed to $20$\% the resulting maximal growth rates are $\lambda \approx 0.2$,
corresponding to a division time of $t_\mathrm{D} = \log(2)/\lambda \approx 3.5h$, within the same order of magnitude
as the fastest known division times for cyanobacteria (we note that simulations are run using nitrate as only nitrogen source). 
The impact of the quota protein fraction on growth (as niche adaptive proteins, NAP) was previously
also discussed in~\citep{Burnap2015}. \label{fig:GrowthQuota}
}
\end{figure}

\clearpage
\subsection{Supplementary Figure S7}
\vspace{3cm} 
\begin{figure}[h!]
    \centering
    \includegraphics[width=.45\textwidth]{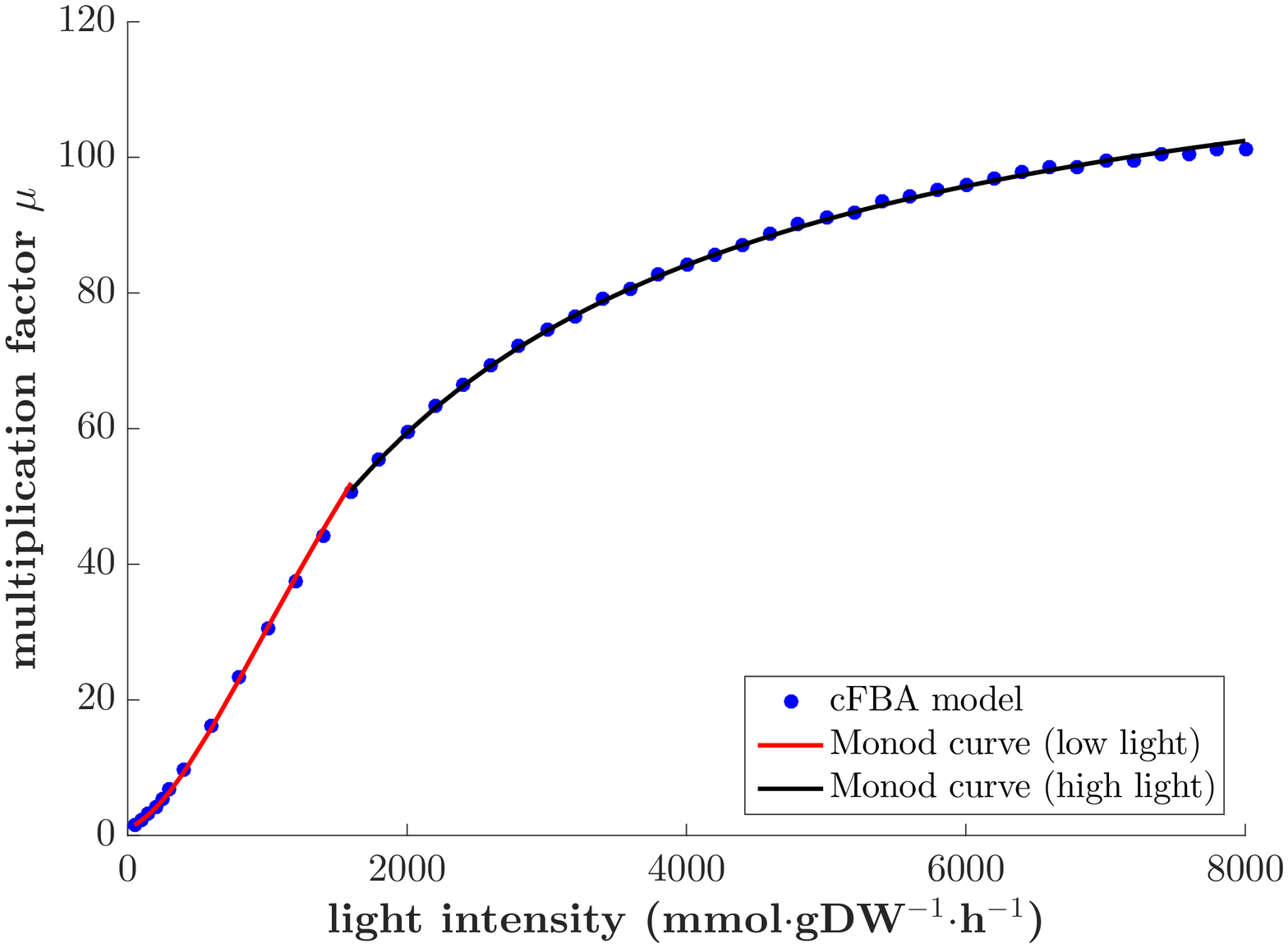}
    \caption{
{\bf Balanced growth under constant light when the percentage of noncatalytic proteins is set to 20\% of the proteome.}
Shown is the maximal multiplication factor $\mu$ as a function of light intensity. The maximal growth rate obtained in this case is $\lambda=0.2019$ h$^{-1}$.\label{fig:GrowthQuota20percent}}
\end{figure}

\clearpage
\subsection{Supplementary Figure S8}
\vspace{3cm} 
\begin{figure}[h!]
    \centering
    \includegraphics[width=.5\textwidth]{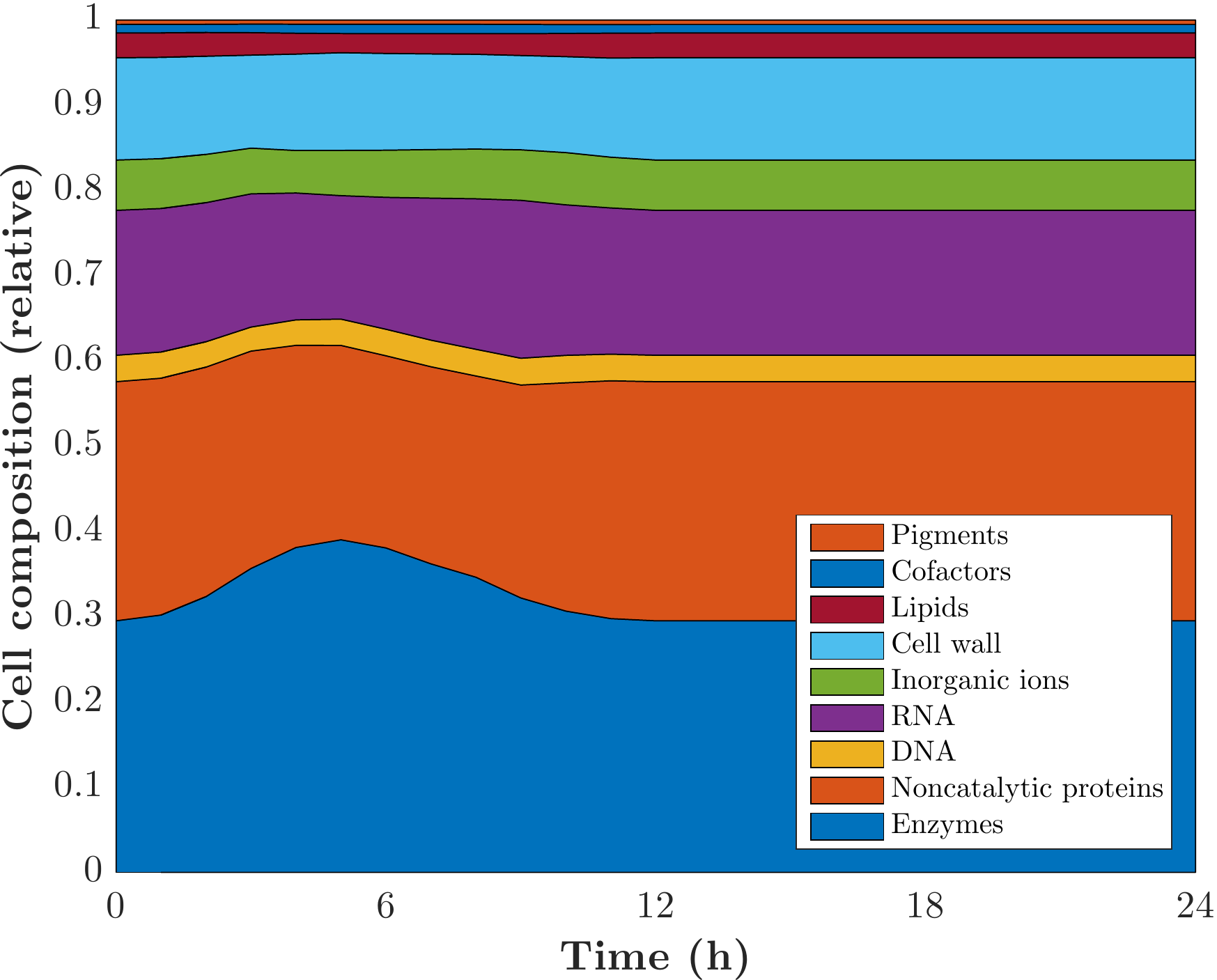}
    \caption{
{\bf Biomass composition over a full diurnal cycle.}
Shown is the relative biomass composition over a full diurnal cycle (12L/12D) as a result of the global resource allocation problem. The simulations were run for a peak light intensity of $600\;\mu$mol photons $\cdot$ s$^{-1}$ $\cdot$ m$^{-2}$.}
\end{figure}

\clearpage
\subsection{Supplementary Figure S9}
\vspace{3cm} 
\begin{figure}[h!]
    \centering
    \begin{minipage}{.4\textwidth}
        \includegraphics[width=.9\textwidth]{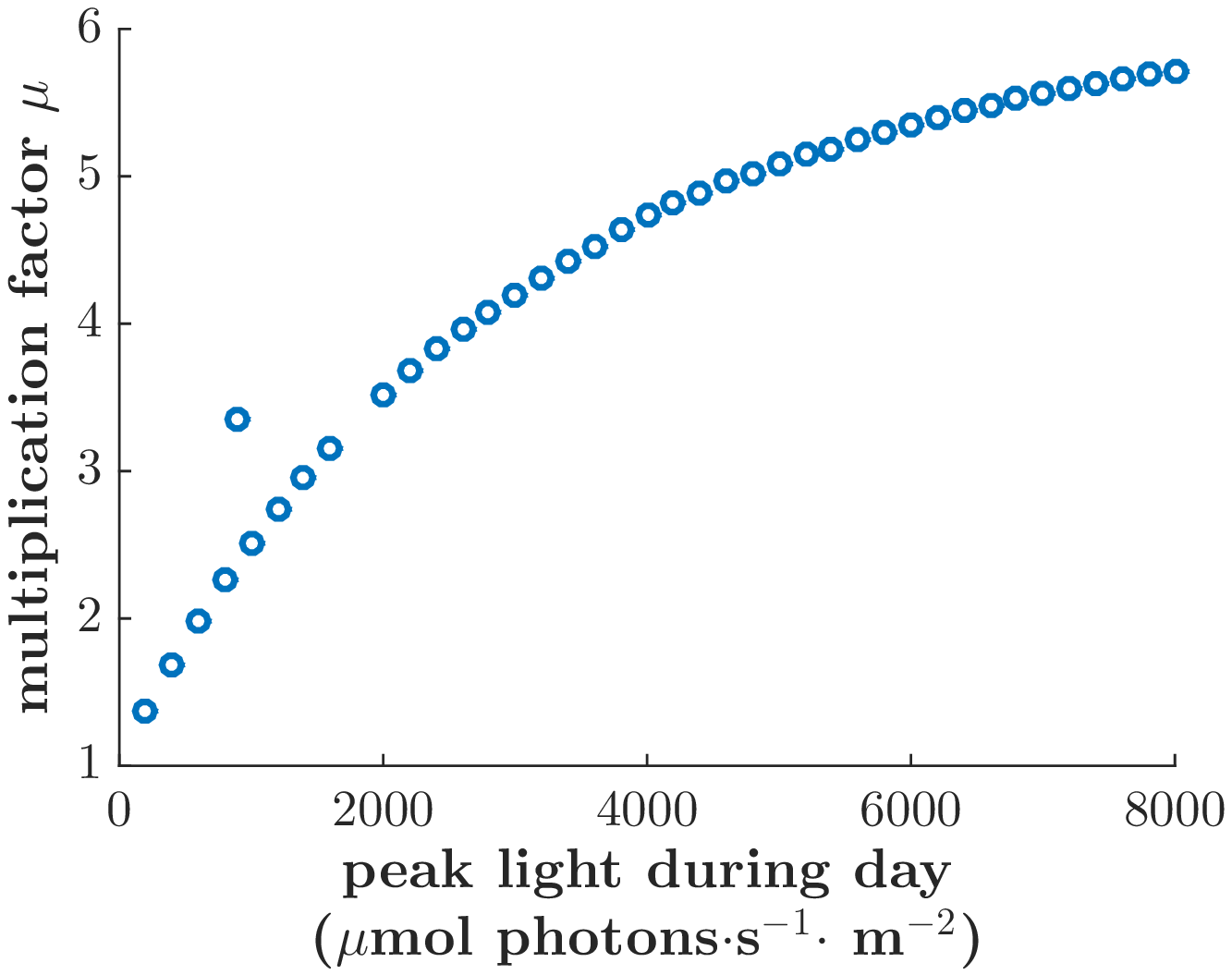}
    \end{minipage}
    \begin{minipage}{.4\textwidth}
        \includegraphics[width=.9\textwidth]{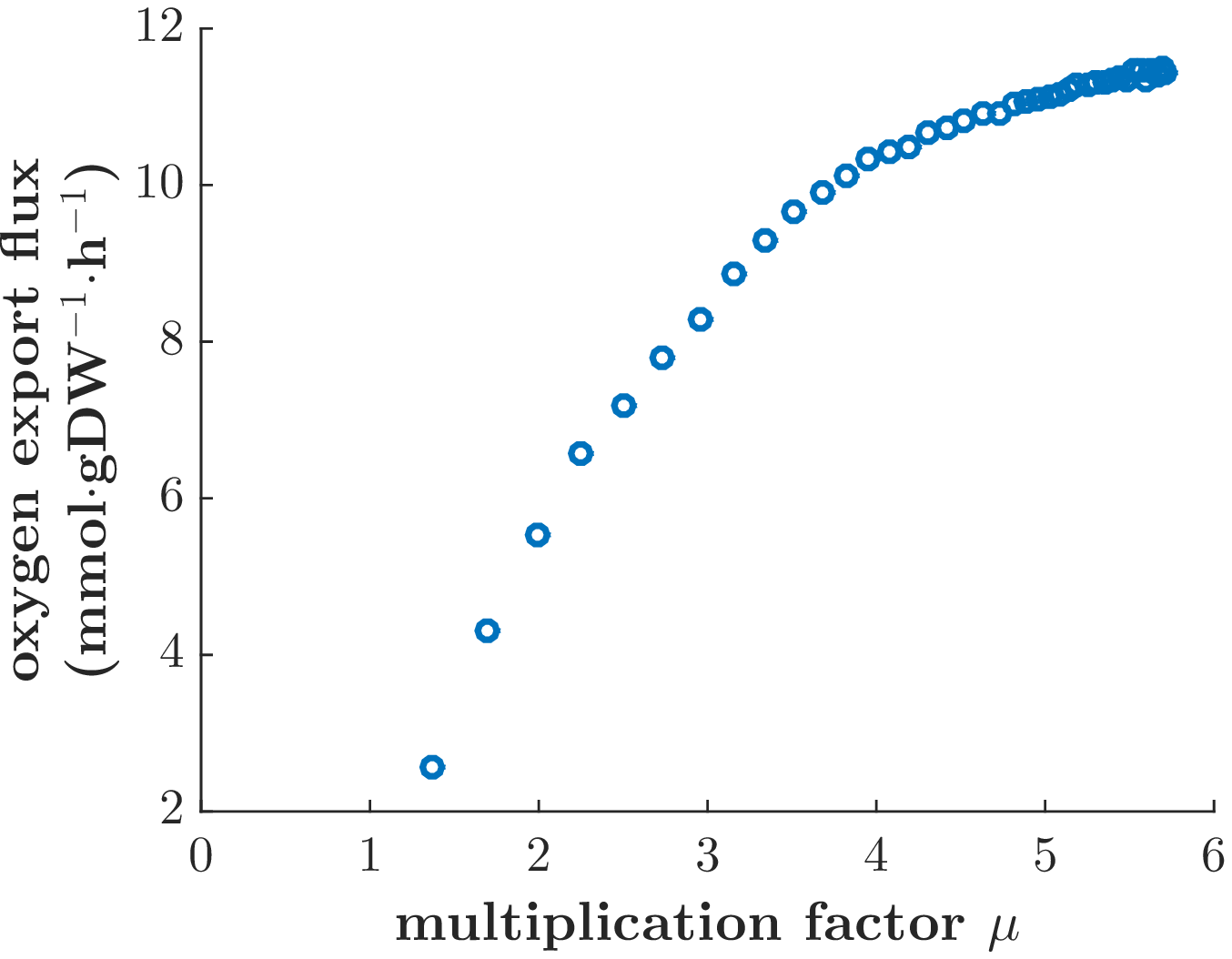}
    \end{minipage}   
 \caption{
{\bf Cellular composition under diurnal light conditions (corresponding to Figure 2 of the main text).}
Left plot: Dependency of the multiplication factor $\mu$ on the peak light intensity. 
Growth increases with increasing light. Light absorption assumes an effective cross section of $\sigma_\mathrm{PSII}=1\mathrm{nm}^2$ for PSII that likely underestimates the true value (see discussion). 
Right plot: Oxygen export flux at noon as a function of the multiplication factor $\mu$. 
The plots are in good qualitative agreement with the results shown for constant light (Figure 2 in the main text). 
\label{fig:CellComp}
}
\end{figure}

\clearpage
\subsection{Supplementary Figure S10}
\vspace{3cm} 
\begin{figure}[h!]
    \centering   
\begin{minipage}{.32\textwidth}
    \includegraphics[width=.9\textwidth]{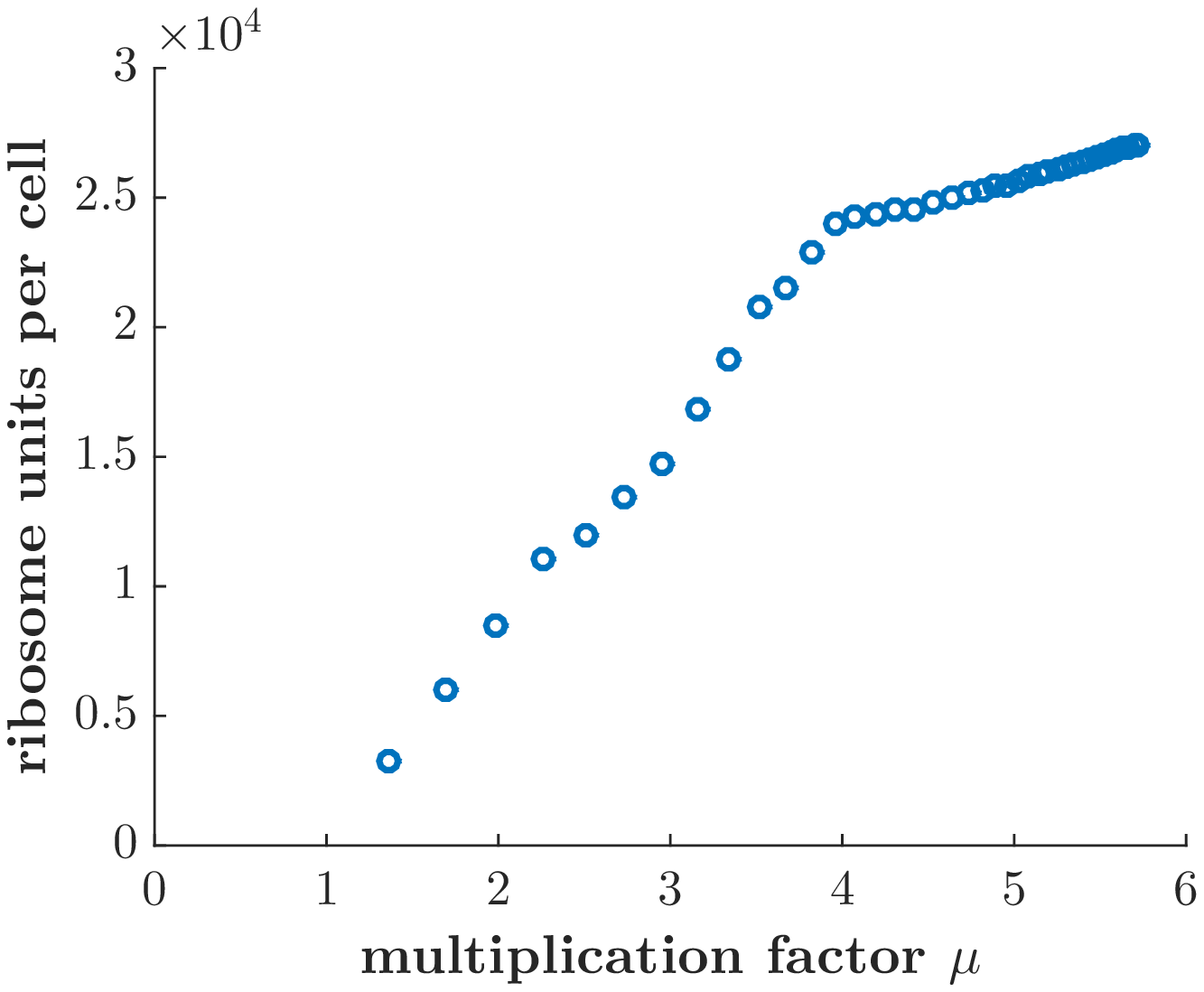}
\end{minipage}
\begin{minipage}{.32\textwidth}
    \includegraphics[width=.9\textwidth]{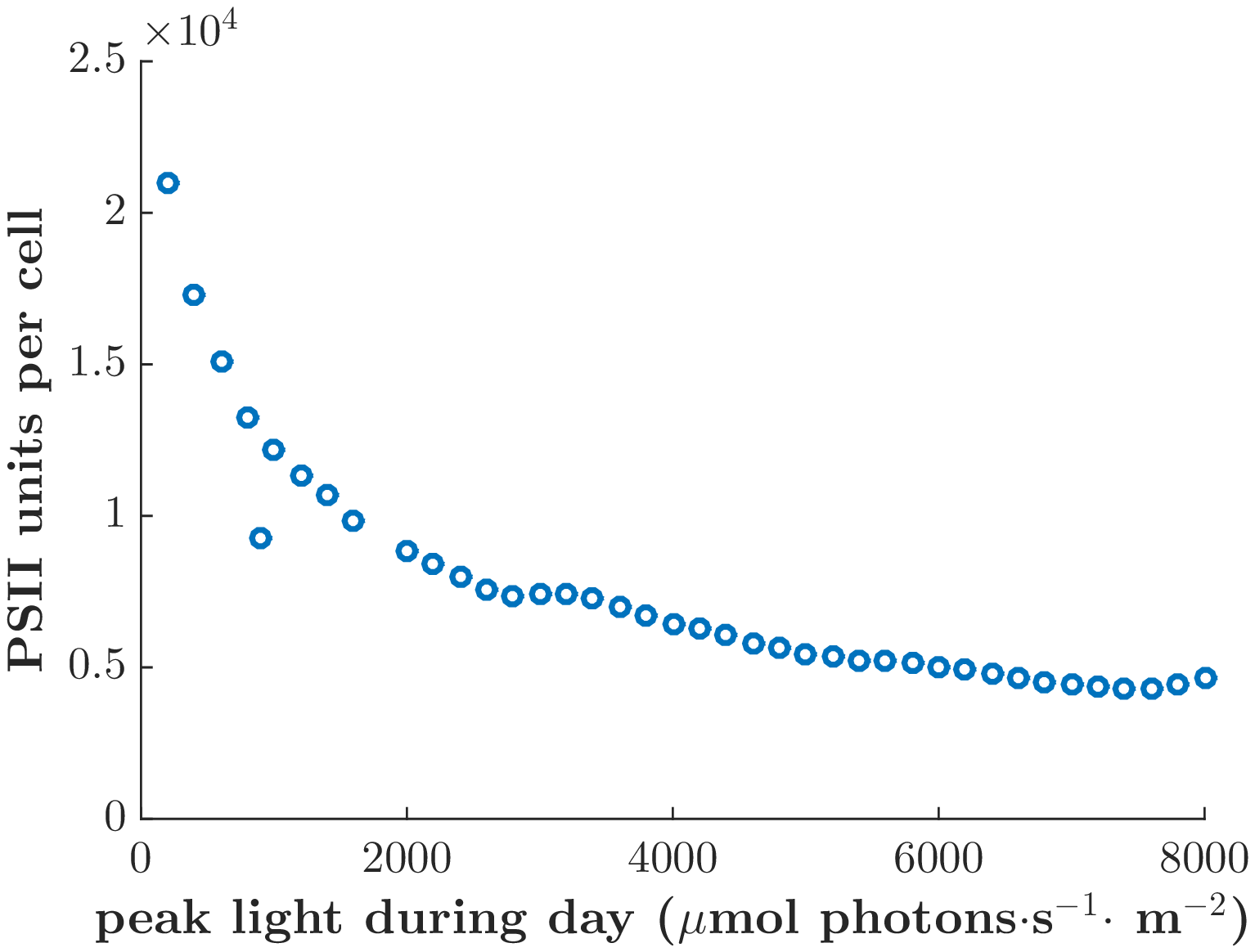}
\end{minipage}
\begin{minipage}{.32\textwidth}
    \includegraphics[width=.9\textwidth]{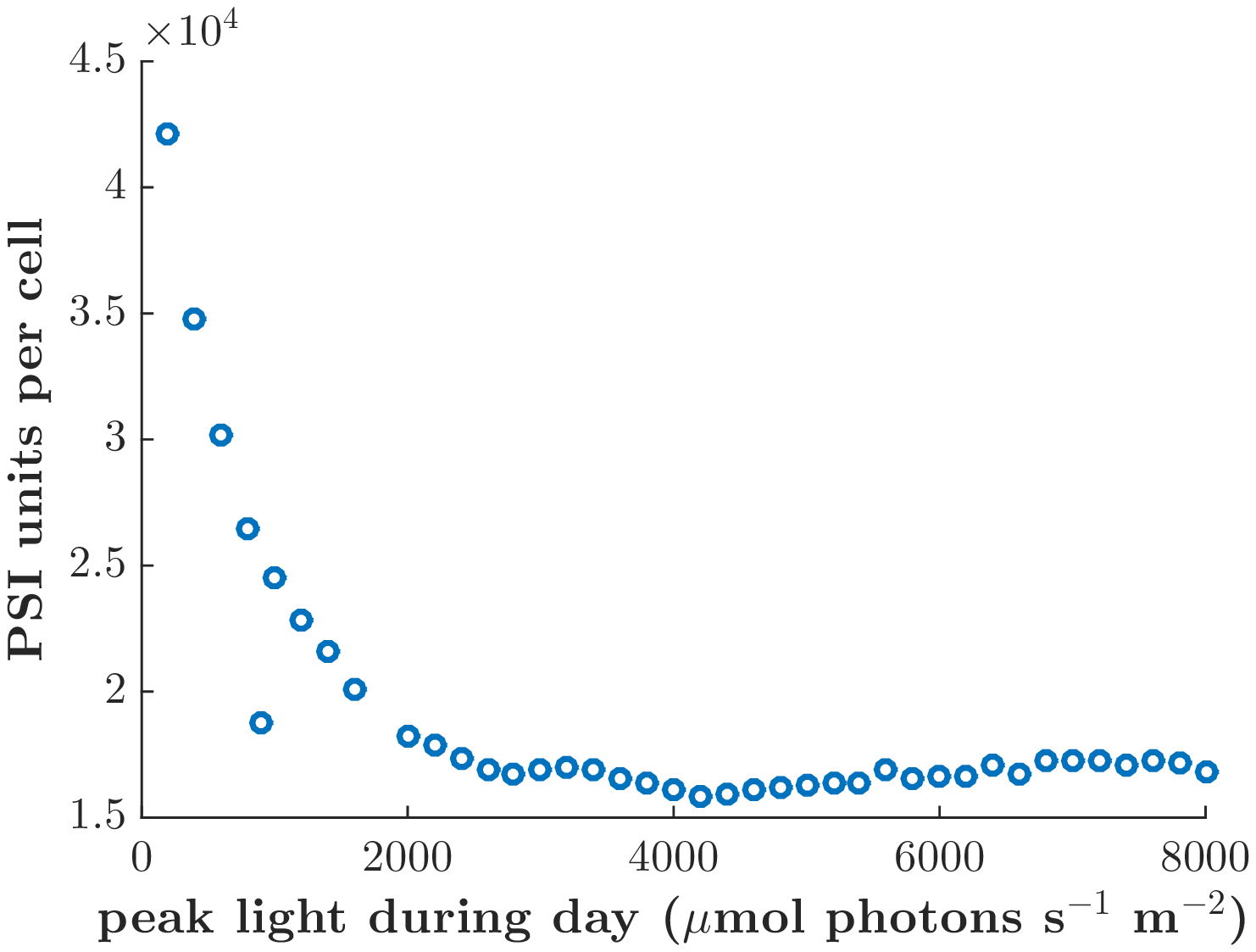}
\end{minipage}
 \caption{
{\bf Cellular composition under diurnal light conditions (corresponding to Figure 2 of the main text).}
Left plot: Dependency of the ribosome content per cell as function of the multiplication factor $\mu$. 
The number was inferred from the number per dry weight, assuming a cell mass of $1.5\mathrm{pg}$.
Middle plot: Dependency of the number of PSII per cell on peak light intensity, assuming an (average) cell mass of $1.5\mathrm{pg}$. 
Right plot: Dependency of the number of PSI per cell on peak light intensity, assuming an (average) mass of $1.5\mathrm{pg}$. 
}
\end{figure}

\clearpage
\subsection{Supplementary Figure S11}
\vspace{3cm} 
\begin{figure}[h!]
    \centering
    \includegraphics[width=.8\textwidth]{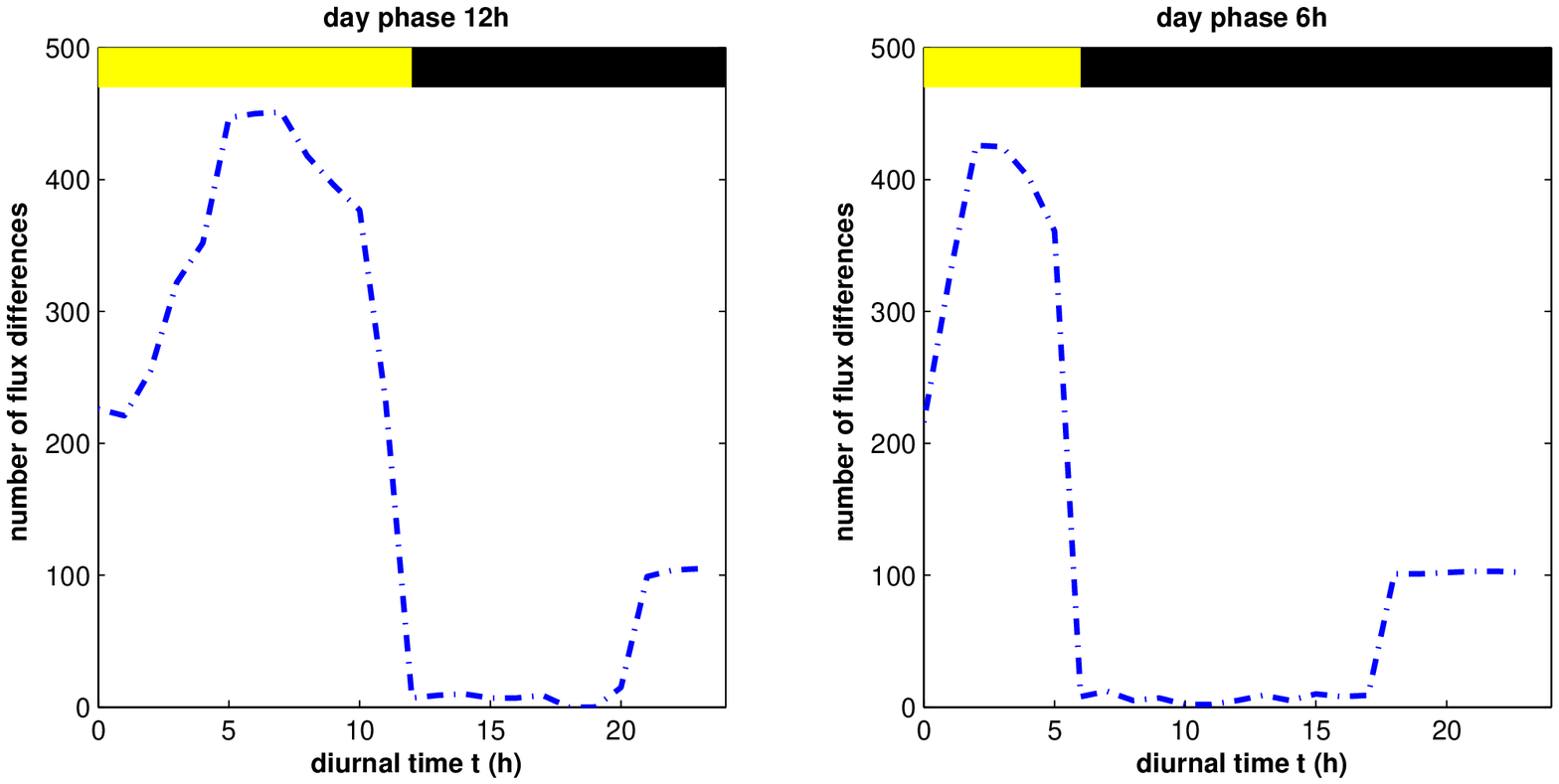}
    \caption{
{\bf Variability of glycogen requirements (corresponding to Figure 5 of the main text).}
Different glycogen requirements at dusk result from variability in resource allocation. 
Shown is the absolute number of metabolic reactions with different activities between high and low glycogen solutions. 
During the light phase, this number includes general flux variability. 
We observe increased variability shortly before dawn. 
Solutions with high glycogen at dusk exhibit
an increase of metabolic activity shortly before dawn, with no effect on overall growth rate. 
\label{fig:GlycogenVariability}
}
\end{figure}

\clearpage
\subsection{Supplementary Figure S12}
\vspace{3cm} 
\begin{figure}[h!]
    \centering
    \includegraphics[width=.95\textwidth]{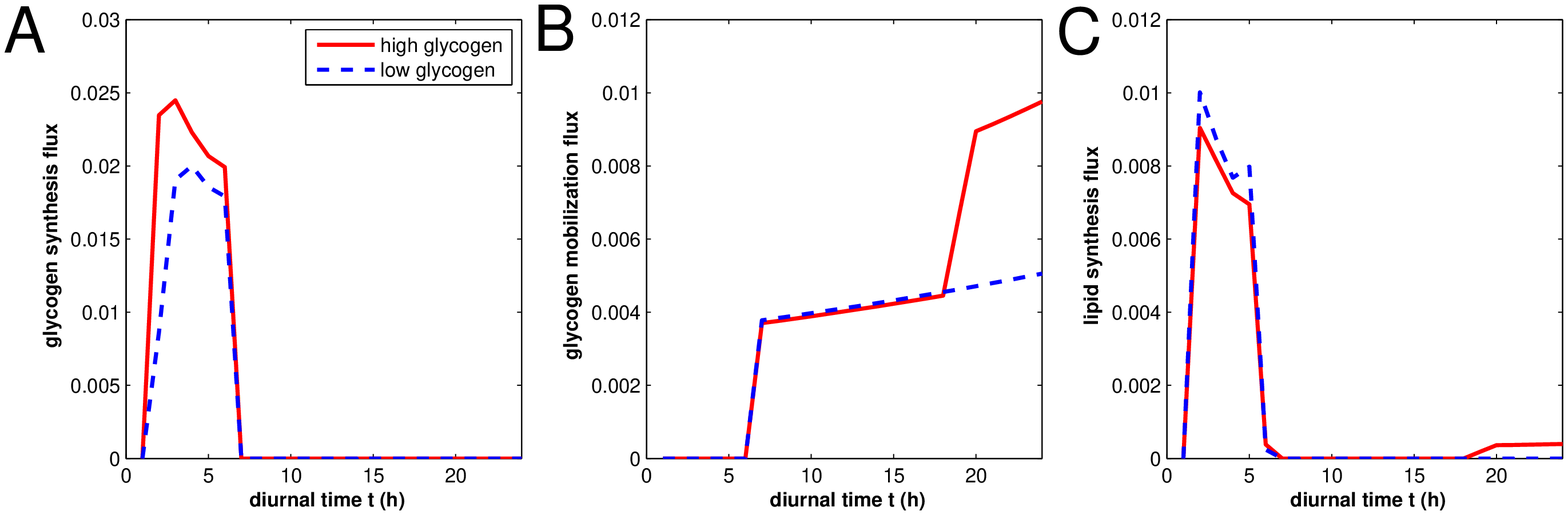}
    \caption{
{\bf Variability of glycogen requirements for a light phase of 6h (corresponding to Figure 5 of the main text).}
Different glycogen requirements at dusk result from variability in resource allocation. 
Common to all day lengths of Figure 5 in the main text is that solutions with high glycogen at dusk correspond to 
an increase of metabolic activity shortly before dawn, with no effect on overall growth rate. 
Specifically, the solutions with more glycogen at dusk synthesize lipids shortly before dawn, requiring less 
total enzyme capacity for lipid synthesis. 
(A) Glycogen synthesis flux.
(B) Glycogen mobilization flux: Solutions with increased glycogen at dusk increase glycogen utilization shortly before dawn.
(C) Lipid synthesis: The additional glycogen is utilized to synthesise lipids. Solutions with high glycogen at dusk 
synthesize less lipids during the day.  
All fluxes are measured relative to gDW in units mmol$\cdot$gDW$^{-1}\cdot$h$^{-1}$. 
We note that flux increase of glycogen mobilization during the night phase is due to a decrease in cell mass.  
\label{fig:6hDAY_GlycogenVariability}
}
\end{figure}

\clearpage
\subsection{Supplementary Figure S13}
\vspace{3cm} 
\begin{figure}[h!]
    \centering
    \includegraphics[width=.95\textwidth]{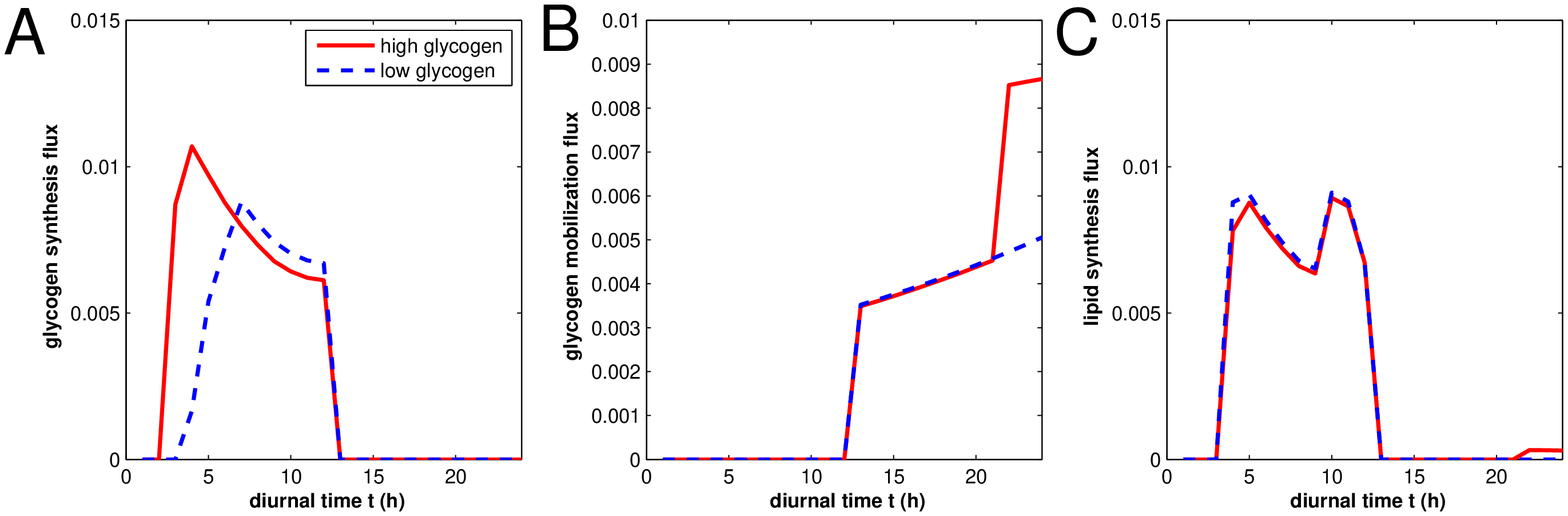}
    \caption{
{\bf Variability of glycogen requirements for light phase of 12h (corresponding to Figure 5 of the main text).}
Same as Figure~\ref{fig:6hDAY_GlycogenVariability} but with a light phase of 12h. 
Solutions are qualitatively identical for different length of the light phase. 
(A) Glycogen synthesis flux.
(B) Glycogen mobilization flux: Solutions with increased glycogen at dusk increase glycogen utilization shortly before dawn.
(C) Lipid synthesis: The additional glycogen is utilized to synthesise lipids. Solutions with high glycogen at dusk 
synthesize less lipids during the day.  
All fluxes are measured in units mmol$\cdot$gDW$^{-1}\cdot$h$^{-1}$.
We note that flux increase of glycogen mobilization during the night phase is due to a decrease in cell mass.  
\label{fig:12hDAY_GlycogenVariability}
}
\end{figure}

\clearpage
\subsection{Supplementary Figure S14}
\vspace{3cm} 
\begin{figure}[h!]
    \centering
    \begin{minipage}{.4\textwidth}
        \includegraphics[width=.95\textwidth]{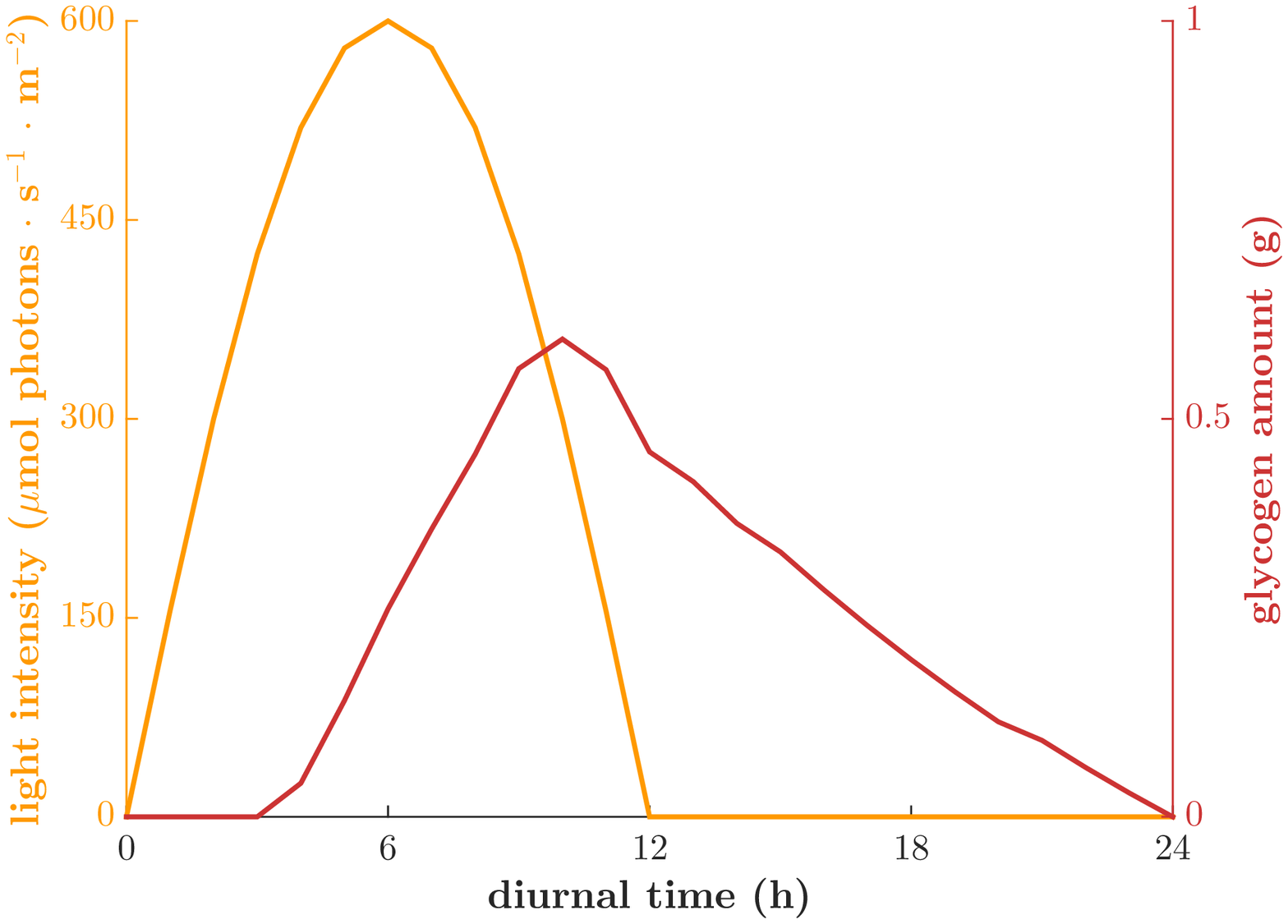}
    \end{minipage}
    \begin{minipage}{.45\textwidth}
        \includegraphics[width=.95\textwidth]{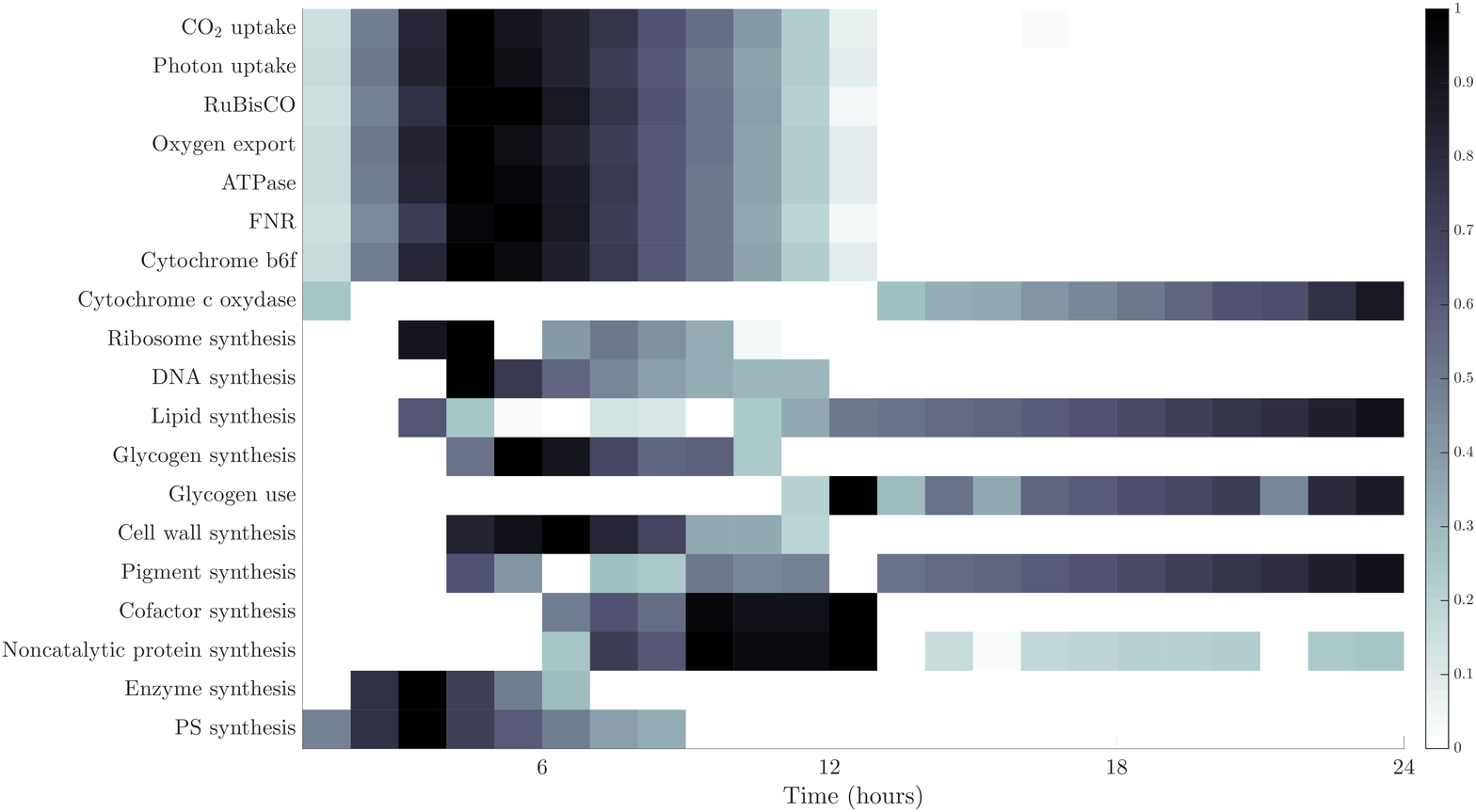}
    \end{minipage}   
 \caption{
{\bf A hypothetical scenario in which the synthesis and breakdown of glycogen requires no enzymatic costs.} 
The reactions catalyzed by the enzymes (ST0001 to ST0006, GS0001 and GS0001$\_$2 in the model file) are
implemented as spontaneous reactions. 
As expected, under these conditions it is energetically more favourable to utilize synthesis
reactions also during the night phase (thereby lowering capacity requirements during the light phase) at the expense of increased glycogen storage. 
Left plot: Timing and dynamics of glycogen accumulation over the day (corresponding to Figure 4A of main text).  
Right plot: metabolic activity of key synthesis reactions. Time courses are color-coded and 
are normalized to the unit interval. 
In particular, lipid synthesis is relegated to the night phase.  
\label{fig:GlycogenFree}
}
\end{figure}

\clearpage
\section{Units, Abbreviations and Notation}
\subsection{List of Abbreviations and Notation}
\begin{itemize}
    \item $\mathbb{R}$ denotes the set of real numbers.
    \item $S$ denotes the stoichiometric matrix of a metabolic network.
    \item $S_i^j$ denotes the stoichiometric coefficient of compound $i$ in reaction $j$.
    \item $\IntMet$ denotes the set of internal metabolites.
    \item $\Enz$ denotes the set of enzymes.
    \item $\rib$ denotes the ribosome.
    \item $\Quota$ denotes the set of quota metabolites (DNA, RNA, cell wall, pigments, nonmetabolic proteins, lipids, cofactors and vitamins, ions).
    \item $\glyc$ denotes glycogen.
    \item $\IntRxn$ denotes the set of metabolic reactions.
    \item $\ERxn$ denotes the set of enzyme production reactions.    
    \item $\ExcRxn$ denotes the set of exchange (transport) reactions.
    \item $\QRxn$ denotes the set of reactions that produce quota metabolites.
    \item $\Irr$ denotes the set of irreversible reactions.
    \item $\v\in\mathbb{R}^{\ExcRxn\cup\IntRxn\cup\ERxn\cup\QRxn}$ denotes the vector of reaction rates (fluxes).
    \item $\c\in\mathbb{R}^{\Enz\cup\Quota\cup\glyc}$ denotes the vector of concentrations.
    \item $\kcat_i$ denotes the turnover rate of reaction $i$.
    \item $MW_i$ denotes the molecular weigh of enzyme $i$ in kDa.
    \item $\mathcal{V}_{j}$ denotes the set of reactions that are catalysed by enzyme $j$.
\end{itemize}
\newpage
\subsection{Units}
\begin{table}[h!]
    \centering
    \begin{tabular}{l p{6cm} r} \hline
        Model instance&Description&Unit\\ \hline \hline
        $\kcat$&Turnover rates&h$^{-1}$\\
        $MW_i$&Molecular weight of enzyme $i$&kDa\\
        $\v_{\IntRxn\cup\ExcRxn\cup\ERxn}$&Metabolic and exchange fluxes&mmol$\cdot$gDW$^{-1}\cdot$h$^{-1}$\\
        $\v_{\QRxn}$&Quota production fluxes&gDW$^{-1}\cdot$h$^{-1}$\\
        $\c_\Quota$&Quota concentrations&g$\cdot$gDW$^{-1}$\\
        $\c_\Enz$&Enzyme concentrations&mmol$\cdot$gDW$^{-1}$\\
        $\c_\glyc$&Glycogen concentration&g$\cdot$gDW$^{-1}$\\
        $\dot\c_\Quota$&Quota concentrations derivatives&g$\cdot$gDW$^{-1}\cdot$h$^{-1}$\\
        $\dot\c_\Enz$&Enzyme concentrations derivatives&mmol$\cdot$gDW$^{-1}\cdot$h$^{-1}$\\
        $\dot\c_\glyc$&Glycogen concentration derivative&g$\cdot$gDW$^{-1}\cdot$h$^{-1}$\\\hline
    \end{tabular}

\end{table}

\clearpage
\section{Supplementary Tables}

\begin{longtable}{|l l r r|}\hline
        Ribosomal protein & Gene & Stoichiometry & Molecular weight (Da)\\\hline
        \multicolumn{4}{|c|}{Small ribosomal subunit}\\\hline
        S1&Synpcc7942\_0694&1&34591\\
        S2&Synpcc7942\_2530&1&28400\\
        S3&Synpcc7942\_2226&1&27718\\
        S4&Synpcc7942\_1487&1&23206\\
        S5&Synpcc7942\_2216&1&19330\\
        S6&Synpcc7942\_0012&1&12346\\
        S7&Synpcc7942\_0886&1&17756\\
        S8&Synpcc7942\_2219&1&14682\\
        S9&Synpcc7942\_2205&1&14937\\
        S10&Synpcc7942\_0883&1&12179\\
        S11&Synpcc7942\_2210&1&13712\\
        S12&Synpcc7942\_0887&1&13991\\
        S13&Synpcc7942\_2211&1&13979\\
        S14&Synpcc7942\_0446&1&11752\\
        S15&Synpcc7942\_2299&1&10308\\
        S16&Synpcc7942\_1772&1&9555\\
        S17&Synpcc7942\_2223&1&9347\\
        S18&Synpcc7942\_1123&1&8307\\
        S19&Synpcc7942\_2228&1&10238\\
        S20&Synpcc7942\_1520&1&10892\\
        S21&Synpcc7942\_1774&1&7029\\\hline
\caption{Ribosome composition (Small ribosomal subunit) of \textit{S. elongatus} 7942.}\label{tbl:ribosome_comp1}
\end{longtable}

\clearpage
\begin{longtable}{|l l r r|}\hline
        Ribosomal protein & Gene & Stoichiometry & Molecular weight (Da)\\\hline
        \multicolumn{4}{|c|}{Large ribosomal subunit}\\\hline
        L1&Synpcc7942\_0633&1&25855\\
        L2&Synpcc7942\_2229&1&31706\\
        L3&Synpcc7942\_2232&1&22452\\
        L4&Synpcc7942\_2231&1&23203\\
        L5&Synpcc7942\_2220&1&20016\\
        L6&Synpcc7942\_2218&1&19198\\
        L7/L12&Synpcc7942\_0631&2x (2-3)=4-6&13151\\
        L9&Synpcc7942\_2559&1&16679\\
        L10&Synpcc7942\_0632&1&18810\\
        L11&Synpcc7942\_0634&1&14903\\
        L13&Synpcc7942\_2206&1&17072\\
        L14&Synpcc7942\_2222&1&13318\\
        L15&Synpcc7942\_2215&1&15289\\
        L16&Synpcc7942\_2225&1&16139\\
        L17&Synpcc7942\_2208&1&13262\\
        L18&Synpcc7942\_2217&1&13047\\
        L19&Synpcc7942\_2541&1&13444\\
        L20&Synpcc7942\_1277&1&13316\\
        L21&Synpcc7942\_1219&1&13870\\
        L22&Synpcc7942\_2227&1&13253\\
        L23&Synpcc7942\_2230&1&11148\\
        L24&Synpcc7942\_2221&1&12449\\
        L27&Synpcc7942\_1220&1&9227\\
        L28&Synpcc7942\_0042&1&9119\\
        L29&Synpcc7942\_2224&1&7650\\
        L31&Synpcc7942\_2204&1&8799\\
        L32&Synpcc7942\_0997&1&6534\\
        L33&Synpcc7942\_1122&1&7372\\
        L34&Synpcc7942\_1614&1&5230\\
        L35&Synpcc7942\_1278&1&7840\\
        L36&Synpcc7942\_2212&1&4364\\\hline
\caption{Ribosome composition (Large ribosomal subunit) of \textit{S. elongatus} 7942.}\label{tbl:ribosome_comp}
\end{longtable}

\begin{longtable}{|l l r r|}\hline
        Ribosomal protein & Gene & Stoichiometry & Molecular weight (Da)\\\hline
        \multicolumn{4}{|c|}{Ribosomal RNA}\\\hline
        16S&Synpcc7942\_R0004&1&\\
        5S&Synpcc7942\_R0006&1&\\
        23S&Synpcc7942\_R0005&1&\\\hline
\caption{Ribosome composition (ribosomal RNA) of \textit{S. elongatus} 7942.}\label{tbl:ribosome_comp3}
\end{longtable}

\clearpage

\begin{table}[H]
    \centering
    \begin{tabular}{|l| l l r|}\hline
        Enzyme&Gene&Gene name&Stoichiometry\\\hline
        \multirow{12}{*}{Photosystem I monomer}&Synpcc7942\_2049&PsaA&1\\
        &Synpcc7942\_2048&PsaB&1\\
        &Synpcc7942\_0535&PsaC&1\\
        &Synpcc7942\_1002&PsaD&1\\
        &Synpcc7942\_1322&PsaE&1\\
        &Synpcc7942\_1250&PsaF&1\\
        &Synpcc7942\_2343&PsaI&1\\
        &Synpcc7942\_1249&PsaJ&1\\
        &Synpcc7942\_0407&PsaK1 (PsaX)&1\\
        &Synpcc7942\_0920&(PsaK2) (PsaX)&1\\
        &Synpcc7942\_2342&PsaL&1\\
        &Synpcc7942\_1912a&PsaM&1\\
        &C05306\_cyt&Chlorophyll a&96\\
        &C02094\_cyt&$\mu$-carotene&22\\
        &C02059\_cyt&Phylloquinone&2\\\hline
    \end{tabular}
    \caption{Gene composition of the photosystem I monomer of \textit{S. elongatus} 7942.}\label{tbl:photosys_comp}
\end{table}

\begin{table}[H]
    \centering
    \begin{tabular}{|l| l l r|}\hline
        \multirow{27}{*}{Photosystem II monomer}&Synpcc7942\_0424 (A)&&1\\
        &Synpcc7942\_0893 (A)&&1\\
        &Synpcc7942\_1389 (A)&&1\\
        &Synpcc7942\_0655 (D1)&PsbA (D1)&1\\
        &Synpcc7942\_1637 (D2)&PsbD (D2)&1\\
        &Synpcc7942\_0697&PsbB&1\\
        &Synpcc7942\_0656&PsbC&1\\
        &Synpcc7942\_1177&PsbE&1\\
        &Synpcc7942\_1176&PsbF&1\\
        &Synpcc7942\_0225&PsbH&1\\
        &Synpcc7942\_1705&PsbI&1\\
        &Synpcc7942\_1174&PsbJ&1\\
        &Synpcc7942\_0456&PsbK&1\\
        &Synpcc7942\_1175&PsbL&1\\
        &Synpcc7942\_0699&PsbM&1\\
        &Synpcc7942\_0224&PsbN&1\\
        &Synpcc7942\_0294&PsbO&1\\
        &Synpcc7942\_1038&PspP&1\\
        &Synpcc7942\_0696&PsbT&1\\
        &Synpcc7942\_1882&PsbU&1\\
        &Synpcc7942\_2010&PsbV&1\\
        &Synpcc7942\_2016&PsbX&1\\
        &Synpcc7942\_1692&PsbY&1\\
        &Synpcc7942\_2245&PsbZ&1\\
        &Synpcc7942\_0343&Psb27&1\\
        &Synpcc7942\_1679&Psb28 (W)&1\\
        &Synpcc7942\_2478&Psb28-2 (W)&1\\
        &C05306\_cyt&Chlorophyll a&35\\
        &C02094\_cyt&$\mu$-carotene&11\\\hline
    \end{tabular}
    \caption{Gene composition of the photosystem II monomer of \textit{S. elongatus} 7942.}
\end{table}

\clearpage

\begin{table}[h]
    \centering
    \begin{tabular}{|l| l l r|}\hline
        Enzyme&Gene&Gene name&Stoichiometry\\\hline
        \multirow{30}{*}{NADPH dehydrogenase}&\textbf{Core (State M):}&&\\
        &Synpcc7942\_1343&NdhA&1\\
        &Synpcc7942\_1415&NdhB&1\\
        &Synpcc7942\_1180&NdhC&1\\
        &Synpcc7942\_1346&NdhE&1\\
        &Synpcc7942\_1345&NdhG&1\\
        &Synpcc7942\_1743&NdhH&1\\
        &Synpcc7942\_1344&NdhI&1\\
        &Synpcc7942\_1182&NdhJ&1\\
        &Synpcc7942\_1181&NdhK&1\\
        &Synpcc7942\_0413&NdhL&1\\
        &Synpcc7942\_1982&NdhM&1\\
        &Synpcc7942\_2234&NdhN&1\\
        &Synpcc7942\_2177&NdhO&1\\
        &\textbf{Variable:}&&\\     
        &&State L:&\\
        &Synpcc7942\_1976&NdhD1&1\\
        &Synpcc7942\_1977&NdhF1&1\\
        &&State L’:&\\
        &Synpcc7942\_1439&NdhD2&1\\
        &Synpcc7942\_1977&NdhF1&1\\
        &&State MS:&\\
        &Synpcc7942\_2092&NdhD3&1\\
        &Synpcc7942\_2091&NdhF3&1\\
        &Synpcc7942\_2093&CupA&1\\
        &Synpcc7942\_2094&CupS&1\\
        &&State MS’:&\\
        &Synpcc7942\_0609&NdhD4&1\\
        &Synpcc7942\_0309&NdhF4&1\\
        &Synpcc7942\_0308&CupB&1\\\hline
        \multirow{12}{*}{Cytochrome b6f}&Synpcc7942\_1232&PetC&2\\
        &Synpcc7942\_2426&PetM&2\\
        &Synpcc7942\_0978&PetH (FNR)&2\\
        &Synpcc7942\_1231&PetA&2\\
        &Synpcc7942\_2331&PetB&2\\
        &Synpcc7942\_2332&PetD&2\\
        &Synpcc7942\_0475&PetN&2\\
        &Synpcc7942\_1479&PetG&2\\
        &Synpcc7942\_0113&PetL&2\\
        &Cytochrome:&&\\
        &Synpcc7942\_0239&cyt f&1\\
        &Synpcc7942\_1630&cyt A (Pet J)&1\\
        &Synpcc7942\_2542&Cyt c6-2&1\\\hline
    \end{tabular}
    \caption{Gene composition of NHD I and Cytochrome b6f.}\label{tbl:NDHCytb6f}
\end{table}

\clearpage

\begin{table}[h]
    \centering
    \begin{tabular}{|p{1.8cm}|p{3.1cm} c c r|}\hline
        Subunit&Description&Gene or&Gene&Stoichiometry\\
        &&metabolite&name&\\\hline
        \multirow{5}{*}{Core}&\multirow{3}{*}{$24\cdot(\alpha \mathrm{AP}+\mu\mathrm{AP})$}&Phycocyanobilin&&48\\
        &&Synpcc7942\_0327&ApcA&24\\
        &&Synpcc7942\_0326&ApcB&24\\\cline{2-5}
        &$6\cdot \mathrm{Lc}$&Synpcc7942\_0325&ApcC&6\\
        &$2\cdot \mathrm{Lcm}$&Synpcc7942\_0328&ApcE&2\\\hline
        \multirow{4}{*}{\parbox{1.8cm}{Rods (length 1)}}&\multirow{3}{*}{$+36\cdot(\alpha\mathrm{PC}+\mu\mathrm{PC})$}&Phycocyanobilin&&+108\\
        &&Synpcc7942\_1048&CpcA&+36\\
        &&Synpcc7942\_1047&CpcB&+36\\\cline{2-5}
        &$+6\cdot\mathrm{Lcr}$&Synpcc7942\_2030&CpcG&+6\\\hline
        \multirow{4}{*}{\parbox{1.8cm}{Rods (length 2)}}&\multirow{3}{*}{$+36\cdot(\alpha\mathrm{PC}+\mu\mathrm{PC})$}&Phycocyanobilin&&+108\\
        &&Synpcc7942\_1048&CpcA&+36\\
        &&Synpcc7942\_1047&CpcB&+36\\\cline{2-5}
        &$+6\cdot\mathrm{Lr}$&Synpcc7942\_1049&CpcC&+6\\\hline
        \multirow{4}{*}{\parbox{1.8cm}{Rods (length 3)}}&\multirow{3}{*}{$+36\cdot(\alpha\mathrm{PC}+\mu\mathrm{PC})$}&Phycocyanobilin&&+108\\
        &&Synpcc7942\_1048&CpcA&+36\\
        &&Synpcc7942\_1047&CpcB&+36\\\cline{2-5}
        &$+6\cdot\mathrm{Lr}$&Synpcc7942\_1049&CpcC&+6\\\hline
    \end{tabular}
    \caption{Gene composition of phycobilisomes in \textit{S. elongatus} 7942. The following abbreviations are used: $\alpha$ PC - Phycocyanin alpha subunit, $\mu$ PC - Phycocyanin beta subunit, $\alpha$ AP - Allophycocyanin alpha subunit, $\mu$ AP - Allophycocyanin beta subunit, Lcr - core-rod linker protein, Lr - rod-rod linker protein, Lc - core-core linker protein, Lcm - core-thylakoid membrane linker protein.}\label{tbl:PBS_comp}
\end{table}

\clearpage

\begin{table}[h]
    \centering
    \begin{tabular}{|l| l l r|}\hline
        Enzyme&Gene&Gene name&Stoichiometry\\\hline
        \multirow{9}{*}{ATPase}&Synpcc7942\_0331&A&1\\
        &Synpcc7942\_0332&C&10-15\\
        &Synpcc7942\_0333&B’&1-2\\
        &Synpcc7942\_0334&B&1-2\\
        &Synpcc7942\_0335&delta&1\\
        &Synpcc7942\_0336&alpha&3\\
        &Synpcc7942\_0337&gamma&1\\
        &Synpcc7942\_2315&beta&3\\
        &Synpcc7942\_2316&epsilon&1\\\hline
        \multirow{3}{*}{Cytochrome c oxidase}&Synpcc7942\_2603&sub I&1\\
        &Synpcc7942\_2602&subII&1\\
        &Synpcc7942\_2604&sub III&1\\\hline
        \multirow{3}{*}{Succinate dehydrogenase}&Synpcc7942\_0314&SdhC&1\\
        &Synpcc7942\_0641&SdhA (Flavoprotein)&1\\
        &Synpcc7942\_1533&SdhB (Iron protein)&1\\\hline
    \end{tabular}
    \caption{Gene composition of ATPase, Cytochrome c oxidase, and succinate dehydrogenase.}\label{tbl:resp_comp}
\end{table}

\clearpage

\begin{table}[h]
    \centering
    \begin{tabular}{|l r r|}\hline
        quota metabolite & fraction (g/gDW)& Amount (pg/cell)\\\hline
        Nonmetabolic proteins&0.357&0.5712\\
        DNA&0.031&0.0496\\
        RNA&0.17&0.272\\
        Cell wall&0.059&0.0944\\
        Lipids&0.12&0.192\\
        Cofactors and vitamins&0.029&0.0464\\
        Ions&0.01&0.016\\
        Pigments&0.0244&0.03904\\\hline        
    \end{tabular}
    \caption{Fractions of the dry weight of a cell that correspond to the quota metabolites as well as absolute amounts of the quota metabolites in one cell, assuming a dry weight of $1.5$ pg/cell. These amounts are used as initial quota amounts in our model.}\label{tbl:initial_quota}
\end{table}

\end{document}